\input harvmac.tex


\lref\PSbook{A.N.~Parshin, I.R.~Shafarevich (Eds.), ``Number Theory II,''
Springer-Verlag, 1992.}

\lref\Shimurabook{G.~Shimura, Y.~Taniyama, ``Complex Multiplication
of Abelian Varieties and Its Applications to Number Theory,''
Japan Math. Soc. 1961.}

\lref\Langbook{S.~Lang (Ed.), ``Number Theory III,''
Springer-Verlag, 1991.}

\lref\Borevicbook{Z.~Borevic, I.~Shafarevich, ``Number theory,''
1985}

\lref\Langgeombook{S.~Lang, ``Fundamentals of Diophantine Geometry,''
Springer-Verlag, 1983.}


\lref\Shimizu{A.~Shimizu, ``On Complex Tori with Many Endomorphisms,''
Tsukuba J.Math. {\bf 8} (1984) 297.}

\lref\Zahrin{F.~Oort, Yu.~Zahrin, ``Endomorphism Algebras of
Complex Tori,'' Math. Ann. {\bf 303} (1995) 11.}

\lref\Nikulin{V.~Nikulin, ``Integral symmetric bilinear forms
and some of their applications,'' Math. Izv. {\bf 14} (1980) 103.}

\lref\Dolgachev{I.~Dolgachev, ``Integral quadratic forms:
applications to algebraic geometry,'' Sem. Bourbaki no.611 (1982) 251.}

\lref\Sloane{J.H.~Conway, N.J.A.~Sloane, ``Sphere Packings, Lattices,
and Codes,'' Springer-Verlag, 1993.}

\lref\Mumford{D.~Mumford, ``A note on Shimura's paper
{\it Discontinuous Groups and Abelian Varieties},''
Math. Ann. {\bf 181} (1969) 345.}

\lref\PShaf{I.~Pjateckii-Shapiro, I.R.~Shafarevich, ``The Arithmetic
of K3 Surfaces,'' Proc. Steklov Inst. Math. {\bf 132} (1973) 45.}

\lref\Borcea{C.~Borcea, ``Calabi-Yau Threefolds and Complex
Multiplication,'' in {\it Essays on Mirror Manifolds,} S.-T.~Yau ed.,
International Press, 1992.}

\lref\Andre{Y.~Andr\'e, ``G-functions and Geometry,''
Aspects of Mathematics, Vol. {\bf E13}, Vieweg, Braunshweig, 1989;
Y.~Andr\'e, ``Distribution des points CM sur les sous-vari\'et\'es
de modules de vari\'et\'es ab\'eliennes, 1997.}

\lref\Oort{F.~Oort, ``Canonical Lifts and Dense Sets of CM-points,''
Arithmetic Geometry, Proc. Cortona symposium 1994, F. Catanese, ed.,
Symposia Math., Vol. XXXVII, Cambridge Univ. Press, 1997, 228.}

\lref\Shioda{T.~Shioda, ``What is known about the Hodge conjecture?''
in {\it Algebraic Varieties and Analytic Varieties},
Adv. Studies Pure Math. {\bf 1} (1983) 55;
T.~Shioda, ``Geometry of Fermat Varieties,''
in {\it Number Theory Related to Fermat's Last Theorem},
Progress in Math. {\bf 26} (1982) 45.}

\lref\Deligne{P.~Deligne, ``Local Behavior of Hodge Structures at
Infinity,'' AMS/IP Studies in Adv. Math. {\bf 1} (1997) 683.}

\lref\MDeligne{P.~Deligne (Notes by J.~Milne), ``Hodge Cycles
on Abelian Varieties,'' Lecture Notes in Math. {\bf 900} (1982) 9.}

\lref\Coleman{R.~Coleman, ``Torsion Points on Curves,''
In Galois representations and arithmetic algebraic geometry
(Y.~Ihara ed.), Adv. Studies Pure Math. {\bf 12} (1987) 235.}

\lref\JNoot{J.~de Jong, R.~Noot, ``Jacobians with Complex
Multiplication,'' in {\it Arithmetic Algebraic Geometry}
(G.~van der Geer, F.~Oort, J.~Steenbrink eds.), Birkh\"auser, 1991.}


\lref\LianYau{B.H.~Lian, S.-T.~Yau, ``Arithmetic Properties of Mirror Map
and Quantum Coupling,'' Commun. Math. Phys. {\bf 176} (1996) 163.}

\lref\MillerMoore{S.~D.~Miller and G.~Moore,
``Landau-Siegel zeroes and black hole entropy,'' arXiv:hep-th/9903267.}

\lref\Moore{G.~Moore, ``Arithmetic and attractors,'' arXiv:hep-th/9807087.}

\lref\Mooreshort{G.~Moore, ``Attractors and arithmetic,'' arXiv:hep-th/9807056.}

\lref\Candelas{P.~Candelas, X.~de la Ossa and F.~Rodriguez-Villegas,
``Calabi-Yau manifolds over finite fields. I,'' arXiv:hep-th/0012233.}

\lref\Schimmrigk{R.~Schimmrigk, ``Arithmetic of Calabi-Yau varieties
and rational conformal field  theory,'' arXiv:hep-th/0111226.}

\lref\Gannon{T.~Gannon, ``Monstrous Moonshine and the Classification
of CFT,'' math.QA/ 9909080.}

\lref\Manin{Y.~Manin, M.~Marcolli, ``Holography Principle and
Arithmetic of Algebraic Curves,'' hep-th/0201036.}


\lref\fs{D.~Friedan, Z.~Qiu, S.~Shenker, ``Conformal Invariance,
Unitarity, and Two-Dimen\-sional Critical Exponents,''
Phys. Rev. Lett. {\bf 52} (1984) 1575.}

\lref\pz{A.A.~Belavin, A.M.~Polyakov, A.B.~Zamolodchikov,
``Infinite Conformal Symmetry in Two-Dimensional Quantum Field
Theory,'' Nucl. Phys. {\bf B241} (1984) 333.}

\lref\verlinde{E.~Verlinde, ``Fusion Rules and Modular Transformations
in 2D Conformal Field Theory,'' Nucl. Phys. {\bf B300} (1988) 360.}

\lref\mooresei{G.~Moore, N.~Seiberg,
``Polynomial Equations for Rational Conformal Field Theories,''
Phys. Lett. {\bf B212} (1988) 451;
``Classical and Quantum Conformal Field Theory,''
Commun. Math. Phys. {\bf 123} (1989) 177;
``Taming The Conformal Zoo,'' Phys. Lett. {\bf 220} (1989) 422.}

\lref\Ishibashi{N.~Ishibashi, ``The Boundary And Crosscap States
In Conformal Field Theories,'' Mod. Phys. Lett. {\bf A4} (1989) 161.}

\lref\Cardy{J.~L.~Cardy, ``Boundary Conditions, Fusion Rules And
The Verlinde Formula,'' Nucl.\ Phys.\ B {\bf 324}, 581 (1989).}

\lref\FSclassalg{J.~Fuchs and C.~Schweigert,
``A classifying algebra for boundary conditions,''
Phys.\ Lett.\ B {\bf 414}, 251 (1997) [arXiv:hep-th/9708141].}

\lref\FSgeneral{J.~Fuchs and C.~Schweigert,
``Branes: From free fields to general backgrounds,''
Nucl.\ Phys.\ B {\bf 530}, 99 (1998) [arXiv:hep-th/9712257].}

\lref\FSone{J.~Fuchs and C.~Schweigert,
``Symmetry breaking boundaries. I: General theory,''
Nucl.\ Phys.\ B {\bf 558}, 419 (1999) [arXiv:hep-th/9902132].}

\lref\FStwo{J.~Fuchs and C.~Schweigert,
``Symmetry breaking boundaries. II: More structures, examples,''
Nucl.\ Phys.\ B {\bf 568}, 543 (2000) [arXiv:hep-th/9908025].}

\lref\RS{A.~Recknagel and V.~Schomerus, ``D-branes in Gepner models,''
Nucl.\ Phys.\ B {\bf 531}, 185 (1998) [arXiv:hep-th/9712186].}

\lref\DVV{R.~Dijkgraaf, E.~Verlinde, H.~Verlinde,
``On Moduli Spaces of Conformal Field Theories with $c \ge 1$,''
Proc. of 1987 Copenhagen Conference {\it Perspectives in String Theory}.}

\lref\Cumrun{C.~Vafa, ``Quantum Symmetries of String Vacua,''
Mod. Phys. Lett. {\bf A4} (1989) 1615.}

\lref\Wilczek{A.~D.~Shapere and F.~Wilczek, ``Selfdual Models
With Theta Terms,'' Nucl.\ Phys.\ B {\bf 320}, 669 (1989).}

\lref\Gutperle{M.~Gutperle and Y.~Satoh,
``D-branes in Gepner models and supersymmetry,''
Nucl.\ Phys.\ B {\bf 543}, 73 (1999) [arXiv:hep-th/9808080].}

\lref\GaberdielR{M.~R.~Gaberdiel and A.~Recknagel,
``Conformal boundary states for free bosons and fermions,''
JHEP {\bf 0111}, 016 (2001) [arXiv:hep-th/0108238].}

\lref\MMS{J.~Maldacena, G.~Moore, N.~Seiberg,
``Geometrical interpretation of D-branes in gauged WZW models,''
JHEP {\bf 0107} (2001) 046.}

\lref\Cappelli{A.~Cappelli, G.~D'Appollonio, ``Boundary States
of $c=1$ and $c=3/2$ Rational Conformal Field Theories,''
JHEP {\bf 0202:039} (2002), hep-th/0201173.}

\lref\OOY{H.~Ooguri, Y.~Oz, Z.~Yin, ``D-Branes on Calabi-Yau
Spaces and Their Mirrors", Nucl.Phys. {\bf B477} (1996) 407.}

\lref\Candelasfst{P.~Candelas, X.~de la Ossa, P.~Green, L.~Parkes,
``A Pair of Calabi-Yau Manifolds as an Exactly Soluble
Superconformal Theory,'' Nucl. Phys. {\bf B359} (1991) 21.}

\lref\IHV{K.~Hori, A.~Iqbal, C.~Vafa,
``D-Branes and Mirror Symmetry,'' hep-th/0005247.}

\lref\Wendland{K.~Wendland,
``Moduli Spaces of Unitary Conformal Field Theories,''
PhD Thesis, September 2000.}

\lref\Tani{S.~Mizoguchi, T.~Tani, ``Wound D-Branes in Gepner Models,''
Nucl. Phys. {\bf B611} (2001) 253.}

\lref\Shamit{S.~Kachru, M.~Schulz, S.~Trivedi, ``Moduli
Stabilization from Fluxes in a Simple IIB Orientifold,'' hep-th/0201028;
see also a talk of S.~Kachru at Strings 2002 Conference,
http://www.damtp.cam.ac.uk/strings02/avt/kachru}


\let\includefigures=\iftrue
\newfam\black
\includefigures
\input epsf
\def\figin{\epsfcheck\figin}\def\figins{\epsfcheck\figins}
\def\epsfcheck{\ifx\epsfbox\UnDeFiNeD
\message{(NO epsf.tex, FIGURES WILL BE IGNORED)}
\gdef\figin##1{\vskip2in}\gdef\figins##1{\hskip.5in}
\else\message{(FIGURES WILL BE INCLUDED)}%
\gdef\figin##1{##1}\gdef\figins##1{##1}\fi}
\def\DefWarn#1{}
\def\figinsert{\goodbreak\midinsert}
\def\ifig#1#2#3{\DefWarn#1\xdef#1{fig.~\the\figno}
\writedef{#1\leftbracket fig.\noexpand~\the\figno}%
\figinsert\figin{\centerline{#3}}\medskip\centerline{\vbox{\baselineskip12pt
\advance\hsize by -1truein\noindent\footnotefont{\bf Fig.~\the\figno:} #2}}
\bigskip\endinsert\global\advance\figno by1}
\else
\def\ifig#1#2#3{\xdef#1{fig.~\the\figno}
\writedef{#1\leftbracket fig.\noexpand~\the\figno}%
\global\advance\figno by1}
\fi

\font\cmss=cmss10 \font\cmsss=cmss10 at 7pt

\def\IB{\relax\hbox{$\inbar\kern-.3em{\rm B}$}}
\def\IC{\relax\hbox{$\inbar\kern-.3em{\rm C}$}}
\def\IQ{\relax\hbox{$\inbar\kern-.3em{\rm Q}$}}
\def\ID{\relax\hbox{$\inbar\kern-.3em{\rm D}$}}
\def\IE{\relax\hbox{$\inbar\kern-.3em{\rm E}$}}
\def\IF{\relax\hbox{$\inbar\kern-.3em{\rm F}$}}
\def\IG{\relax\hbox{$\inbar\kern-.3em{\rm G}$}}
\def\IGa{\relax\hbox{${\rm I}\kern-.18em\Gamma$}}
\def\IH{\relax{\rm I\kern-.18em H}}
\def\IK{\relax{\rm I\kern-.18em K}}
\def\IL{\relax{\rm I\kern-.18em L}}
\def\IP{\relax{\rm I\kern-.18em P}}
\def\IR{\relax{\rm I\kern-.18em R}}
\def\Z{\relax\ifmmode\mathchoice
{\hbox{\cmss Z\kern-.4em Z}}{\hbox{\cmss Z\kern-.4em Z}}
{\lower.9pt\hbox{\cmsss Z\kern-.4em Z}}
{\lower1.2pt\hbox{\cmsss Z\kern-.4em Z}}\else{\cmss Z\kern-.4em
Z}\fi}

\def\II{\relax{\rm I\kern-.18em I}}


\def\CF {{\cal F}}

\def\CI {{\cal I}}

\def\CN {{\cal N}}
\def\CO {{\cal O}}

\def\CT {{\cal T}}

\def\CV {{\cal V}}


\def\tilde{\widetilde}
\def\hat{\widehat}
\def\bar{\overline}



\def\inbar{\,\vrule height1.5ex width.4pt depth0pt}

\def\a{\alpha}


\Title{\vbox{\baselineskip12pt
\hbox{hep-th/0203213}
\hbox{ITEP-TH-19/02}
\hbox{HUTP-02/A004}}}
{\vbox{
\centerline{Rational Conformal Field Theories and}
\centerline{Complex Multiplication}
\vskip 4pt
}}
\centerline{Sergei Gukov and Cumrun Vafa}
\medskip
\medskip
\medskip
\medskip
\vskip 8pt
\centerline{\it Jefferson Physical Laboratory}
\centerline{\it Harvard University}
\centerline{\it Cambridge, MA 02138, USA}
\medskip
\medskip
\noindent

We study the geometric interpretation of two dimensional rational
conformal field theories, corresponding to sigma models on Calabi-Yau
manifolds.  We perform a detailed study of RCFT's corresponding
to $T^2$ target and identify the Cardy branes with geometric branes.
The $T^2$'s leading to RCFT's admit ``complex multiplication''
which characterizes Cardy branes as specific D0-branes.
We propose a condition for the conformal sigma model
to be RCFT for arbitrary Calabi-Yau $n$-folds,
which agrees with the known cases.
Together with recent conjectures by mathematicians it appears
that rational conformal theories are not dense in the space of
all conformal theories,  and sometimes appear to be
finite in number for Calabi-Yau $n$-folds for $n>2$.
RCFT's on K3 may be dense.  We speculate about the meaning
of these special points in the moduli spaces of Calabi-Yau
$n$-folds in connection with freezing geometric moduli.

\smallskip
\Date{March 2002}

%
%

\newsec{Introduction}
Two dimensional
``Rational Conformal Field Theories'' (RCFT)
were introduced in \fs\ as a particularly
 nice class of conformal field theories which have more structure and could
be potentially classified.  Moreover it was suggested that perhaps
they may be dense in the space of all conformal theories, and so in this
way one can potentially get a handle on all conformal theories.
RCFT's are characterized by having a symmetry algebra
extending the Virasoro algebra (chiral algebra),
in terms of which the Hilbert space can be decomposed into finite
irreducible representations.  Thus RCFT's generalize
the notion of minimal models introduced in \pz.
Their conjecture motivated a great
deal of work on RCFT's leading in particular to Verlinde algebra
\verlinde\ and the rich structure they encode \mooresei.
Moreover it was shown in \refs{\Cardy,\Ishibashi}\ that
RCFT's naturally lead to boundary states, which in modern terminology
we call D-brane states.

In the post duality era, we have learned the importance of D-branes
in uncovering non-perturbative aspects of string theory.  Thus it is
natural to ask the following question:  We consider strings propagating
on a Calabi-Yau background and we vary the moduli of Calabi-Yau.  At points
on the moduli which correspond to RCFT's we naturally have some
finite number of `special' D-branes.  What do these D-branes correspond to?
How do we interpret their preferred role at those moduli among
the infinitely many allowed D-branes?

Interesting as these question appear, the modern discovery
of S-duality raises a further question:  Is the notion of RCFT
an S-duality invariant concept?  The answer to this is no \Moore.
In fact it could hardly be an S-duality invariant
concept because the S-dual theory may not even correspond to string theory
so there is no notion of 2d conformal theory on the S-dual side.
Even if the S-dual theory is a string theory, one can easily
see that the RCFT's on one side do not correspond to RCFT's
on the dual side.  For example type IIB is self-dual, with
the roles of fundamental and D-strings exchanged;
the rationality of its toroidal compactifications
strongly depends on the $B_{NS}$ but is independent of $B_R$.
However on the S-dual side the roles of $B_{NS}$ and $B_R$ are exchanged.
Thus rationality is not an S-duality invariant concept.
One might thus consider this concept as not being a fundamental concept.
However, it turns out that at least in some cases the concept of
special points on moduli space makes sense even non-perturbatively.
For example the same considerations that apply to rational points on
compactification of strings on $T^2$ lead to singling out
special points on the moduli space of the type IIB string coupling
constant $\tau$ which could potentially have some significance in the
full non-perturbative theory.  Another way such a concept
may remain relevant non-perturbatively is exemplified
by compactification on Calabi-Yau 3-folds.  In this case the
string coupling constant combines with other hypermultiplets
which come from K\"ahler moduli (in the type IIB case) and so the
question of rationality, which seems to split
between K\"ahler and Complex moduli, picks out,
in a non-perturbative sense, some special complex
moduli on the Calabi-Yau.
More generally, at the very least the moduli corresponding to
RCFT's must be somehow special at weak string coupling and
thus must teach us some extra symmetries about the target space
physics.

In this paper we take up the question of RCFT's for
sigma models on Calabi-Yau $n$-folds.  More specifically we
consider the case of complex dimension $1$ in detail and use
it to advance a conjecture about rational points for the more complicated
case of sigma models on Calabi-Yau $n$-folds.

For the case of (complex dimension one)
$T^2$ target space we uncover the extra symmetry
principle for the target space for it to correspond
to a RCFT.  These correspond to tori which
admit {\it Complex Multiplication}.  This means that there is a complex
number $\lambda$ (not real) for which $z\rightarrow \lambda z$
maps $T^2$ to itself.  This is not necessarily an isomorphism, and
is in general a many to one map.  Moreover the corresponding K\"ahler class
has to be a complex multiplier $\lambda$ for this to correspond to a RCFT
with a diagonal modular invariant.  Moreover for each K\"ahler class
there exists a canonical complex multiplication which has the significance
that gives the corresponding Cardy states as D0 branes localized
at preimages of a given point on $T^2$ under this complex multiplication.

In higher dimensions, mathematicians have a generalized
notion of complex multiplication \refs{\Mumford, \PShaf, \Borcea},
which is natural to conjecture
is related to the notion of rationality of conformal theory.
This is basically the statement that the mid-dimensional cohomology
and the associated variation of Hodge structure, leads naturally
to the period matrix of a higher dimensional torus and one asks
whether the associated torus admits a complex multiplication.
In the case of Calabi-Yau threefolds this leads to a complex
torus corresponding to coupling constant matrix of the gauge fields.
In the geometric engineering of $\CN=2$ theories, this is the associates
complex torus encoding the BPS masses of $\CN=2$ electric and magnetic
charge states.  Translated in this way, there is a mathematical
conjecture which suggests that in many cases
there are only a {\it finite} number
of points on moduli where the theory is rational.  This is in
sharp contrast to the case of $T^2$ where the rational points
are dense in the space of all conformal theories.

The organization of this paper is as follows:
In the next section we review a simple example
of $c=1$ RCFT based on a circle, which will help us
to introduce the notations and relevant concepts.
In section 3 we discuss two families of RCFT based
on two-dimensional tori with extra symmetries:
$(a)$ direct product of two circles, and $(b)$ a torus
with $\Z_3$ symmetry. Both examples have been
extensively studied in the literature, and we use them to
illustrate general features of rational conformal field theories.
Section 4 gives a friendly introduction into basics of
imaginary quadratic number fields, which play a central
role in the characterization of rational CFT's.
Following these introductory sections, in section 5
we proceed to the general case of $c=2$ CFT based on
elliptic curve, and formulate the criteria for CFT to be rational
and, further, to be diagonal. Our results allow to
classify such rational conformal field theories.
In section 6 we discuss geometric and arithmetic
interpretation of Cardy states in RCFT based on elliptic curve.
Finally, in section 7 we conjecture a generalization of
these results to Calabi-Yau manifolds of higher dimension.

In different contexts, relation between string theory
and number theory has been discussed previously in
\refs{\LianYau,\Moore,\MillerMoore,\Candelas,\Schimmrigk,\Shamit,\Manin}.


\newsec{Review of a Compact Free Boson}

We start with a review of $c=1$ CFT associated with
a free bosonic field on a circle of radius $R$.
Since this theory was extensively studied in the literature,
and has been nicely reviewed in a number of recent papers,
see {\it e.g.} \refs{\FSgeneral, \MMS, \Cappelli, \GaberdielR},
here we briefly describe only
some aspects that will be relevant in the following sections.
In our conventions $\alpha'=1$,
so that the self-dual radius is $R_{s.d}=1$.

For generic values of the radius $R$, the torus partition
function has the following form:
\eqn\sonepfncn{Z (q, \bar q; R) = {1 \over \vert \eta \vert^2}
\sum_{ (p, \bar p) \in \Gamma^{1,1}}
q^{\half p^2} \bar q^{\half \bar p^2} }
where $q = \exp (2 \pi i \tau)$  and
Dedekind's $\eta$-function is defined as
$$\eta = q^{1/24} \prod_{n=1}^{\infty} (1 - q^n)$$
The partition function $Z (q, \bar q; R)$
is given by a sum over the even, self-dual momentum
lattice $\Gamma^{1,1}$. Explicitly, the left and the right
momenta read:
\eqn\ppbarsone{p = {1 \over \sqrt{2}} \Big( {n \over R} + m R \Big),
\quad
\bar p = {1 \over \sqrt{2}} \Big( {n \over R} - mR \Big) }

By definition, the theory is rational if one can represent
the partition function $Z(q,\bar q)$
as a {\it finite} sum of the form:
\eqn\zrcft{Z (q, \bar q) = \sum_{j,j} M_{j \bar j}
\chi_{j} (q) \bar \chi_{\bar j} (\bar q) }
where $M_{j \bar j} \in \Z_{\ge 0}$ and
$\chi_{i}$ (resp. $\bar \chi_{j}$) are holomorphic
(resp. anti-holomorphic) characters:
$$
\chi_j (q) = {\rm tr}_{\CV_j} q^{L_0 - c/24}
$$
For the toroidal examples $\chi_{i}$ and $\bar \chi_{j}$
are generalized $\theta$-functions with characteristics.

Modular invariance and existence of the unique vacuum
state impose further restrictions on the matrix $M$.
Namely, uniqueness of the vacuum implies:
\eqn\mnotnot{M_{00}=1}
On the other hand, modular invariance of the torus partition
function requires:
\eqn\tandscomm{[M,T]=0, \quad [M,S]=0}
where matrices $S$ and $T$ determine transformation
of the characters under $SL(2,\Z)$:
\eqn\tsdefs{\eqalign{
\chi_i (- 1/\tau) = & \sum_{j} S_{ij} \chi_j (\tau) \cr
\chi_i (\tau + 1) = & \sum_{j} T_{ij} \chi_j (\tau)
}}
The matrix $T$ is diagonal and can be written in terms
of the conformal dimensions $\Delta_i$ and the central charge $c$:
\eqn\tmatrix{T_{ij} = \delta_{ij} e^{2 i \pi (\Delta_i - c/24)}}
There is no such simple general expression for the matrix $S$.
However, in RCFT's where Verlinde algebra is an abelian group algebra
$S_{ij}$ are proportional to roots of unity, which follows from
the fact that $S$ diagonalizes Verlinde algebra.
For example, in the rational $c=1$ CFT of a compact boson, we have:
\eqn\sonesmatrix{ S_{jj'} = {1 \over \sqrt{2N}} e^{- i \pi jj' / N}}
For a given RCFT,
it is an interesting problem to classify integer matrices $M$,
which satisfy the relations \mnotnot\ and \tandscomm,
see \Gannon\ for a review.

Now, let us explain a geometric interpretation of the rationality
condition in a theory of a free compact boson. In other words,
we want to analyze when the exponent set $\CI = \{ i \}$ becomes finite.
{}From the explicit form of the left and right momenta \ppbarsone\
it is clear that this happens when $R^2$ is a rational number:
$$
R^2 = {k \over l}, \quad k,l \in \Z
$$
where $k$ and $l$ are relatively prime integer numbers.
The partition function in this case reads:
$$
Z (q, \bar q) =
\sum_{{ i + j = 0~{\rm mod}~2k \atop i-j=0~{\rm mod}~ 2l}}
\chi_{i} (q) \bar \chi_{j} (\bar q)
$$
It is manifestly invariant under T-duality symmetry,
which among other things inverts the radius $R$ and
exchanges the winding and momentum modes:
$$
\Z_2 \quad : \quad
R \leftrightarrow {1 \over R}, \quad m \leftrightarrow n
$$

The chiral primaries in this theory are labeled by
index $j \in \Z$ mod $2kl$.
All theories with the same value of $N=kl$
have the same fusion ring,
which in this simple case is just the group algebra of:
$$
\Z_{2N}
$$
Therefore, theories with the same chiral algebra,
but different modular invariants correspond
to different ways one can decompose $N$ into
a product of two integers.
In particular, $R^2 = {\rm integer}$
(or $R^2 = 1/{\rm integer}$) correspond to
diagonal modular invariants $M_{j \bar j} = \delta_{j \bar j}$
(or charge conjugation modular invariants):
\eqn\diagonal{Z (q, \bar q)
= \sum_{j \in \CI} \chi_{j} (q)~ \bar \chi_{j} (\bar q)}
%

\subsec{Cardy States}

Let us now discuss D-branes in rational conformal field theories.
Among all consistent boundary states, there is a special
(finite) subset of states, which are invariant under the full
extended chiral algebra.
This distinguished set of states, called Cardy states,
can be systematically constructed in a diagonal RCFT
following the original work of Cardy \Cardy.

Let us recall that Cardy states are
linear combinations of the Ishibashi states obtained by
imposing a modular invariance on the string world-sheet:
\eqn\cardysol{ \vert i \rangle
= \sum_i { S_{ij} \over \sqrt{S_{0j}}} \vert j \rangle \rangle }
In particular, the number of the Cardy states is equal to the number
of the Ishibashi states.

The Ishibashi states, in turn, are defined as generalized
solutions to the gluing conditions \Ishibashi:
\eqn\gluingcond{
\Big(W_n-(-1)^{\Delta_W} \Omega(\bar W_{-n}) \Big) \vert B \rangle \rangle =0}
where $\Omega$ is a gluing automorphism,
and $\Delta_W$ is the conformal dimension of
the chiral algebra operator $W$.
For example, in the case of a free compact boson,
the boundary state should preserve the $U(1)$ current
$J = i \sqrt{2N} \partial X$.

Since the Virasoro generators are quadratic in
the oscillator modes, the Ishibashi boundary condition
\gluingcond\ is solved by:
\eqn\boundarycond{
( \alpha^{\mu}_n - R^{\mu}_{\nu} \tilde \alpha^{\nu}_{-n} )
\vert B \rangle \rangle = 0}
We can also write this condition as:
\eqn\boundarycondx{
\partial X^{\mu} (z)
= R^{\mu}_{\nu} \bar \partial X^{\nu} (\bar z)}
%
Here we restored space-time
indices $\mu, \nu = 1, \ldots, d$, and $R$ is an automorphism
of the chiral algebra,
such that $R \in O(d)$ and $\vert {\rm det} (R) \vert =1$.
In particular, $+1$ eigenvalues of the automorphism correspond
to Neumann boundary conditions, whereas $-1$ eigenvalues
correspond to Dirichlet boundary conditions \FSgeneral.
For this reason, a boundary state corresponding to
an automorphism with $p$ eigenvalues $+1$ is referred to as a $Dp$-brane.

For a given automorphism $R$, the explicit form of the boundary
state $\vert B \rangle \rangle$ satisfying \boundarycond\ is given by:
\eqn\bstate{\vert B \rangle \rangle
= \exp \Big[ - \sum_{n=1}^{\infty} {1 \over n}
\alpha^{\mu}_{-n} R_{\mu \nu} \tilde \alpha^{\nu}_{-n} \Big]
\times \sum
\vert p, \bar p \rangle
}
The sum over momenta in this expression goes over all
the elements in the momentum lattice, which satisfy
the condition \boundarycond.

Let us now come back to the case of the diagonal RCFT corresponding
to a free boson on a circle of radius $\sqrt{N} R_{s.d.}$.
In this case, there are only two choices for
the automorphism, $R=\pm1$, corresponding
to D1-branes and D0-branes, respectively.
In the latter case, we obtain $2N$ Ishibashi states,
the A-states in the notations of \refs{\OOY,\MMS}:
\eqn\soneaishi{
\vert A n , n \rangle \rangle =
\exp \Big[ + \sum_{n=1}^{\infty} {1 \over n}
\alpha_{-n} \tilde \alpha_{-n} \Big]
\times \sum_m
\vert {n \over \sqrt{N}} + m \sqrt{N} ,
{n \over \sqrt{N}} + m \sqrt{N} \rangle
}
where $n \in \Z$ mod $2N$.
For the other automorphism one has only
two Ishibashi B-states:
\eqn\sonebishi{
\vert B n , - n \rangle \rangle =
\exp \Big[ - \sum_{n=1}^{\infty} {1 \over n}
\alpha_{-n} \tilde \alpha_{-n} \Big]
\times \sum
\vert {n \over \sqrt{N}} + m \sqrt{N} ,
- {n \over \sqrt{N}} - m \sqrt{N} \rangle
}
where $n=0$ or $n=N$.

Using the Cardy's formula \cardysol\ and the explicit
form for the elements of matrix $S$, in this model
one finds two boundary states corresponding to D1-branes
with different values ($\pm 1$) of the Wilson line,
and $2N$ D0-branes located at the equidistant positions
on the circle, see Figure 1:
%
%
$$
\# ({\rm Dp-branes})~= \cases{2N & $p=0$ \cr 2 & $p=1$}
$$
These Cardy states will be our basic building blocks
in a specific example of $c=2$ RCFT discussed next.

\ifig\pic{Cardy states in rational $c=1$ CFT can be identified
with equally spaced D0-branes (black dots) on a circle.}
{\epsfxsize2.0in\epsfbox{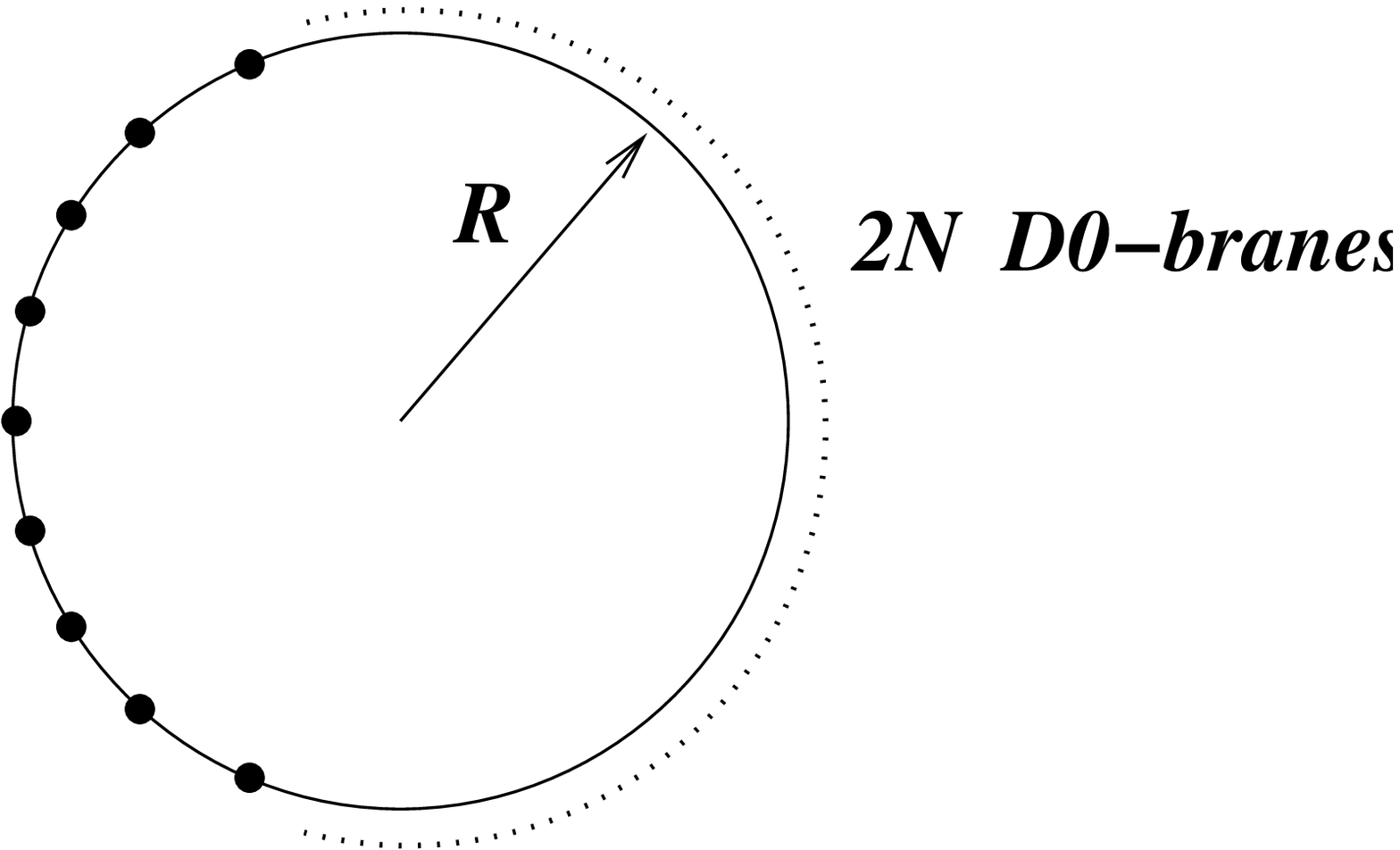}}


\newsec{Simple Examples of $c=2$ RCFT}

Starting with examples in this section, we proceed to our
main subject, namely $c=2$ RCFT based on the elliptic curve $E$.
Specifically, we
analyze the conditions on the complex structure parameter $\tau$
and the complexified K\"ahler modulus $\rho$ under which
the theory becomes rational and has a diagonal partition function \diagonal.
These results will help us build some intuition about
what should happen in the general case that will be discussed
in the following sections.

\subsec{Basic Notations}

To begin, we summarize our conventions and
recall the discrete symmetries of a general RCFT based
on the elliptic curve (see also \DVV\ for a nice exposition).
Throughout the paper we use the following notations
for the real and imaginary parts of $\tau$ and $\rho$:
\eqn\taurhodef{\eqalign{
\tau & = \tau_1 + i \tau_2
= {G_{12} \over G_{11}} + i {\sqrt{{\rm det} G} \over G_{11}} \cr
\rho & = \rho_1 + i \rho_2 = B + i \sqrt{{\rm det} G} }}
where $G_{ij}$ denote components of the (flat) metric on $E$.
Indeed, it is straightforward to check that
\eqn\ttwometric{\eqalign{
ds^2 & = {\rho_2 \over \tau_2} \vert dx + \tau dy \vert^2 = \cr
& = G_{11} dx^2 + 2 G_{12} dx dy + G_{22} dy^2}}

Since we are interested in geometric interpretation
of RCFT, we further assume that:
\eqn\physconstr{\tau_2 > 0, \quad \rho_2 > 0}

RCFT based on the elliptic curve $E$ enjoys
a large group of discrete symmetries:
\eqn\symmetrygroup{
PSL(2,\Z)_{\tau} \times PSL(2,\Z)_{\rho}
\times \Z_2 \times \Z_2 \times \Z_2}
Apart from the last factor, this symmetry group may be viewed
as a group of T-dualities.
In particular, the first two factors in this group act via modular
transformations on $\tau$ and $\rho$, respectively. For example:
$$
PSL(2,\Z)_{\tau} \colon \tau \mapsto {a \tau + b \over c \tau + d}
$$
The $\Z_2$ factors, on the other hand, interchange $\tau$, $\rho$,
and their complex conjugates. Specifically, the first $\Z_2$
acts as:
\eqn\ztwoone{\Z_2 \colon (\tau,\rho) \mapsto (\rho, \tau)}
By analogy with the corresponding symmetry of
higher dimensional varieties, we refer to
this $\Z_2$ as to the mirror transform.

The second $\Z_2$ factor in \symmetrygroup\ is space-time
parity transformation acting as follows:
\eqn\ztwotwo{\Z_2 \colon (\tau,\rho) \mapsto (- \bar \tau, - \bar \rho)}
Finally, the last $\Z_2$ factor in \symmetrygroup\ reverses
world-sheet orientation:
\eqn\ztwothree{\Z_2 \colon (\tau,\rho) \mapsto (\tau, - \bar \rho)}

\subsec{A Product of Two Circles: $E = {\bf S}^1 \times {\bf S}^1$}

Our first simple example of $c=2$ RCFT will be a product
of two $c=1$ rational CFT's corresponding to a product of
two circles of radii:
$$
R_1
= \sqrt{k_1 \over l_1}
\quad {\rm and} \quad
R_2
= \sqrt{k_2 \over l_2}
$$
where $k_i$ and $l_i$ are some (pairwise co-prime) integers.
In these models both $\tau$ and $\rho$ are pure imaginary
(modulo real integer part, which can be set to zero by
$PSL(2,\Z)$ transformations):
$$
\tau = i \sqrt{{k_1 l_2 \over k_2 l_1}},
\quad
\rho = i \sqrt{{k_1 k_2 \over l_1 l_2}}
$$
Note, that both $\tau$ and $\rho$ satisfy quadratic equations
with integer coefficients (and the same discriminant).

Using the analogous result for a single compact boson,
we find that RCFT based ${\bf S}^1 \times {\bf S}^1$ has
a diagonal modular invariant if $l_1=1$ and $l_2=1$,
{\it i.e.} up to modular transformations:
$$
\tau = i \sqrt{{k_1 \over k_2}},
\quad
\rho = i \sqrt{{k_1 k_2}}
$$
Another way to describe these $\tau$ and $\rho$ is to say
that $\tau$ is a solution to the quadratic equation:
\eqn\exonequadr{k_2 \tau^2 + k_1 =0}
while $\rho$ is an integer multiple of $\tau$:
$$
\rho = k_2 \tau
$$
This (strange) relation between $\tau$ and $\rho$ is a precursor
of the general property of the elliptic curve, called complex
multiplication. As we will see in the following sections,
all elliptic curves with this property correspond to a diagonal RCFT,
and the converse is also true.

Since in the present example, the RCFT is a product of two
theories, the Verlinde algebra is just a product:
$$
\Z_{2k_1} \times \Z_{2k_2}
$$
Note, that the total dimension of the chiral ring is equal to $4 k_1 k_2$.
This also gives the number of D0-branes in this theory.
Indeed, boundary states corresponding to D-branes in the $c=2$
RCFT in consideration can be obtained by tensoring the suitable
Cardy states in the two copies of $c=1$ RCFT. Specifically,
a product of D$p_1$ boundary state with D$p_2$ boundary state
gives a D$(p_1+p_2)$-brane boundary state:
$$
\vert D(p_1 + p_2) \rangle =
\vert Dp_1 \rangle \otimes \vert Dp_2 \rangle
$$

If we tensor two A-type Cardy states, we obtain
boundary states corresponding to D0-branes, $4 k_1 k_2$ in number.
The D0-branes are distributed on a torus in a regular lattice of
$2 k_1$ rows and $2 k_2$ columns, as shown in Figure 2.
On the other hand, if we tensor two B-type boundary states,
we get four D2-branes, which cover the entire torus and
differ in the values of Wilson lines they carry. Specifically,
since each D1-brane on a circle carries $\pm 1$ Wilson line,
by tensoring two of them we get four boundary states,
labeled by $(\pm 1, \pm 1)$.

Finally, tensoring A-type Cardy states with B-type Cardy states
gives D1-branes parallel to the sides of the torus.
Namely, if we tensor D0-brane state on the first circle with
a D1-brane state on the second circle, we obtain D1-branes
wrapped on the second circle inside $E = {\bf S}^1 \times {\bf S}^1$.
Since we could take $2k_1$ boundary states corresponding to D0-branes
and there are 2 possible choices for the Wilson line on the D1-brane,
in total we get $2 \times 2 k_1 = 4 k_1$ parallel D1-branes.
Similar arguments show that there are $4k_2$ parallel D1-branes
wrapped on the other basic cycle of the torus, see Figure 2.
Hence, there are $4(k_1+k_2)$ D1-branes in total.

Summarizing, we have:
\eqn\dsquaretorus{
\# ({\rm Dp-branes})~= \cases{4k_1k_2 & $p=0$
\cr 4(k_1 + k_2) & $p=1$ \cr 4 & $p=2$}}

\ifig\pic{Cardy states in rational CFT based on a torus
$E={\bf S}^1 \times {\bf S}^1$.}
{\epsfxsize3.0in\epsfbox{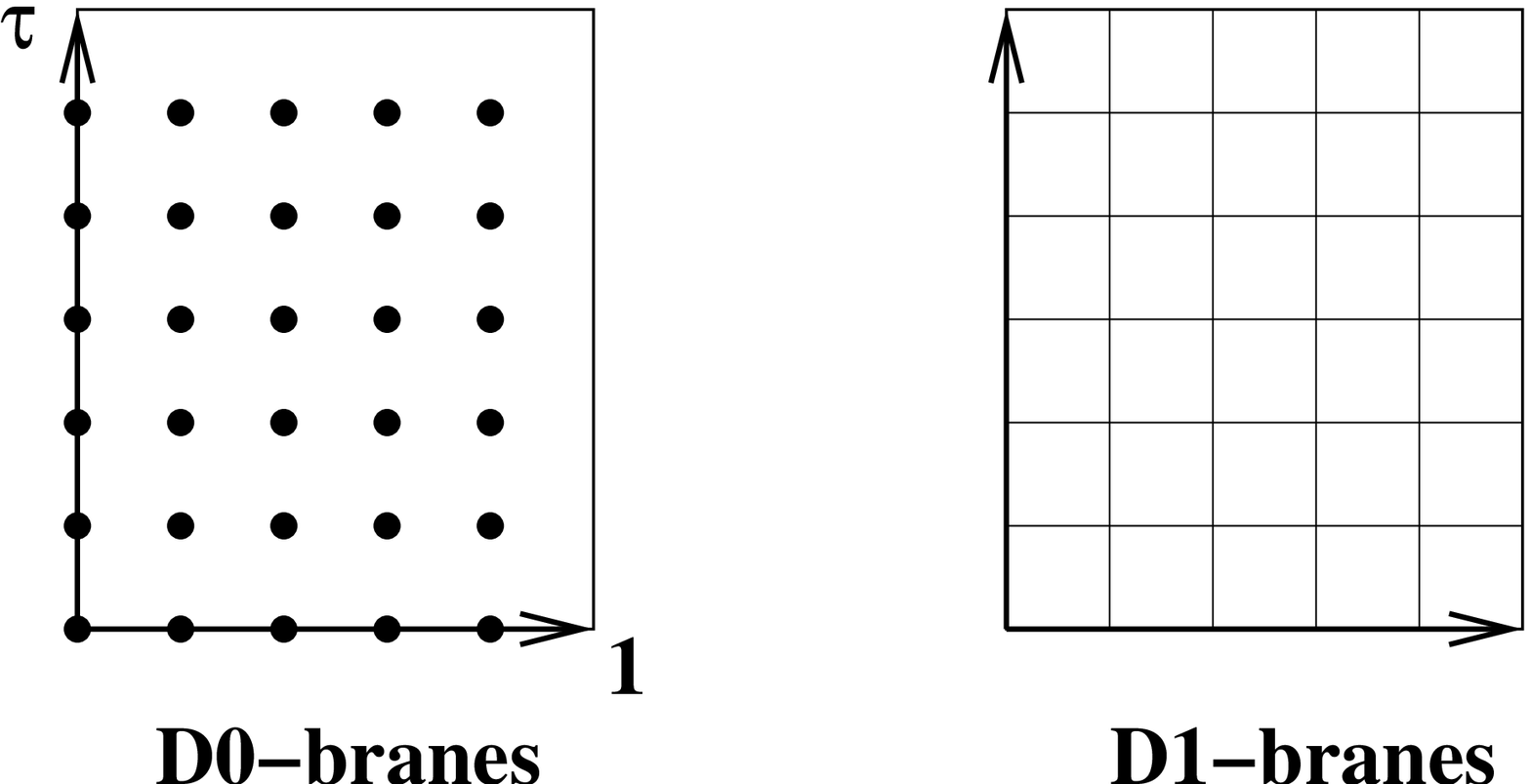}}

\subsec{$SU(3)$ Torus}

There is another simple example of $c=2$ RCFT based on
the torus that has been extensively studied
in the literature \refs{\OOY,\RS,\Gutperle}.
It is $SU(3)$ WZW theory at level 1 corresponding to
elliptic curve $E$ with extra $\Z_3$ symmetry
\Cumrun:
$$
\tau = \rho = \exp(2 \pi i / 3)
$$
As in the previous example, in this case both
$\tau$ and $\rho$ also satisfy a quadratic equation:
%
\eqn\extwoquadr{\tau^2 + \tau + 1 =0}

The dimension of the chiral ring of this theory is
equal to 3, and the Verlinde algebra is just
a group algebra of:
$$
\Z_3
$$
Comparing this with the previous example, one might
expect that in general the fusion rules are given
by (a product of) cyclic groups. In what follows we
will show that this is indeed always the case;
specifically, the Verlinde algebra is a group algebra of:
\eqn\verlindeguess{\Z_{n_1} \times \Z_{n_2}}
This guess includes the special case of a single cyclic group
(as in the present example) when one of the factors is trivial, $n_i =1$.

Cardy states corresponding to various D-branes in this model
have been studied in a number of papers \refs{\OOY,\RS,\Gutperle,\Tani}.
Here, we summarize the result:
\eqn\dsuthreetorus{
\eqalign{
\# ({\rm D0-branes}) &= 3 \cr
\# ({\rm D2-branes}) &= 1
}}
The number of D0-branes is expected to be 3 on general grounds.
In fact, in all theories the number of Cardy states corresponding
to D0-branes should be equal to the dimension of the chiral ring,
which is indeed 3 in the present case.

Since generic elliptic curve $E$ does not reduce to a product
of two circles, at the moment we have nothing to say
about D1-branes. We can just briefly mention that certain boundary
states in this model, studied in \refs{\OOY,\RS,\Gutperle, \Tani},
can be identified with D1-branes wrapped on the shortest cycles
of the torus rotated by $\pi /3$, as illustrated on Figure 3.
In the general discussion below we will find the complete set
of boundary states in this theory.

A little bit more interesting is a result for D2-branes.
Combining it with the result of the previous example,
one might conclude that the number of D2-branes can be at least 1 or 4.
Quite surprisingly, we will show that this a general answer:
in all $c=2$ RCFT's based on elliptic curve, there is either one or
four D2-branes. The number depends on certain arithmetic properties of $E$.

\ifig\pic{D-branes on the $SU(3)$ torus.}
{\epsfxsize3.0in\epsfbox{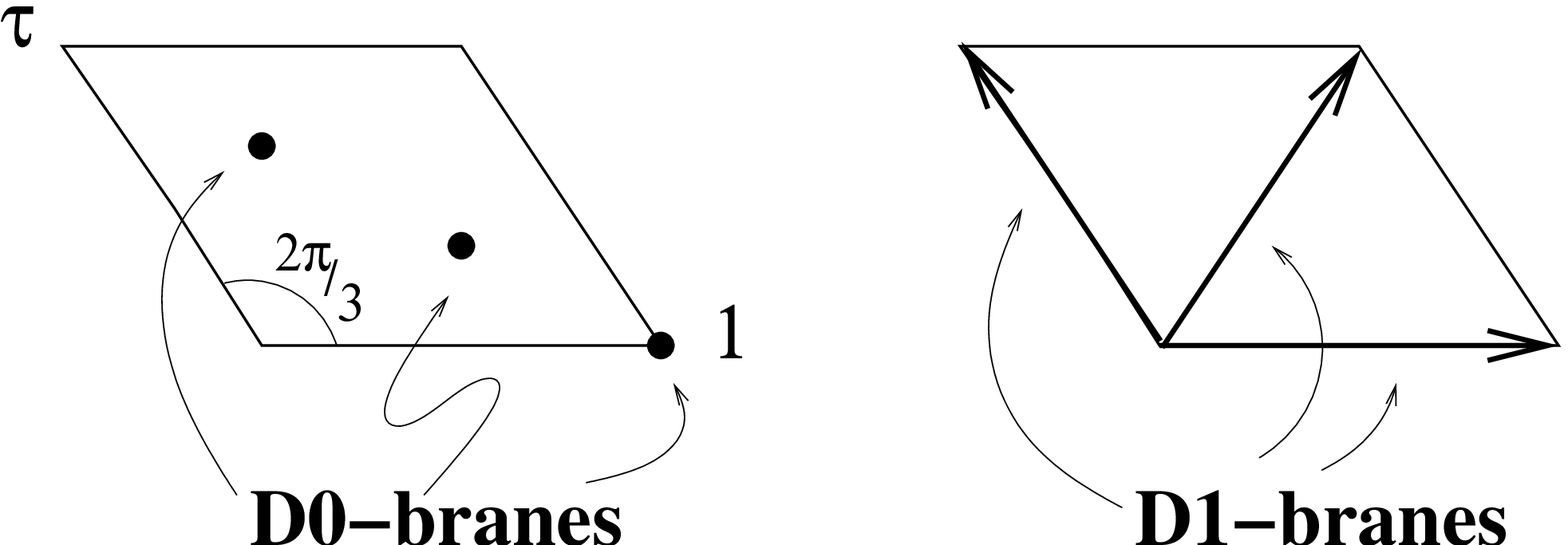}}


\newsec{Imaginary Quadratic Number Fields}

Before we discuss the general case of (rational) conformal
field theory based on the elliptic curve $E$ we need
to introduce a few basic notions from number theory
that will naturally appear in our discussion.
In fact, some of these objects have already entered our
discussion in the implicit form.
For example, in two special cases discussed in the previous
section we have noticed that complex parameter $\tau$
satisfies a quadratic equation of the form,
{\it cf.} \exonequadr\ and \extwoquadr:
\eqn\quadreq{a \tau^2 + b \tau + c = 0}
with relatively prime integer coefficients $a$, $b$, and $c$.
We shall call this quadratic equation a minimal polynomial
for $\tau$, and denote by $D$ its discriminant:
\eqn\disc{D = b^2 - 4 ac}
In all cases relevant to physics $D<0$, so that $\tau$
has a non-zero imaginary part:
\eqn\tausol{\tau = {-b + \sqrt{D} \over 2a}}

Mathematically, it means that $\tau$ is valued
in the {\it imaginary quadratic number field}:
$$
K = \IQ(\sqrt{D})
$$
This particular way of writing the number field $K$
indicates that it can be obtained from the familiar
field of all rational numbers, $\IQ$, by introducing $\sqrt{D}$.
In other words, every number $x$ in $\IQ (\sqrt{D})$
can be written in the form:
$$
x = \alpha + \beta \sqrt{D}
$$
where $\alpha$ and $\beta$ are rational numbers,
$\alpha, \beta \in \IQ$.
Notice, the way we construct the number field
$\IQ(\sqrt{D})$ from rational numbers is very similar
to how one usually defines the field of complex
numbers, $\IC$, supplementing the field of real
numbers, $\IR$, with $\sqrt{-1}$.

In our applications, $\tau$ is not just a number -- it is
a modulus of the elliptic curve:
\eqn\equotient{E = \IC / (\Z \oplus \tau \Z)}
It turns out that
elliptic curves with modular parameter $\tau \in \IQ(\sqrt{D})$
have a nice property called {\it complex multiplication}
\refs{\PSbook,\Shimurabook,\Langbook}.
Elliptic curves with this property enjoy a lot of
wonderful arithmetic and geometric properties.

To explain what complex multiplication means,
let us consider endomorphisms of the elliptic curve $E$,
{\it i.e.} holomorphic maps from $E$ to itself:
$$
\varphi \quad : \quad E \to E
$$
Note, $\varphi$ is a finite degree map (not necessarily degree one).
To describe such maps more explicitly, we can view
the elliptic curve $E$ as a quotient of a complex plane
(parameterized by $z$) by a lattice $\Z \oplus \tau \Z$,
{\it cf.} \equotient.
Then, an endomorphism $\varphi$ simply acts as $z \mapsto \varphi z$.
Since $\Z \oplus \tau \Z$ is a two-dimensional lattice,
we have only have to verify that $\varphi$ maps its generators
to some other other elements in this lattice:
\eqn\cmmap{\eqalign{
\varphi \cdot 1 & = m_1 + n_1 \tau, \cr
\varphi \cdot \tau & = m_2 + n_2 \tau}}
Clearly, any elliptic curve has many trivial endomorphisms
corresponding to multiplication by an integer, $\varphi \in \Z$.
In order to see if there exist any non-trivial endomorphisms,
one can take $\varphi$ from the first equation and substitute
it to the second equation. As a result, one finds a quadratic
equation with integer coefficients of the form \quadreq.
This simple calculation illustrates that elliptic curves
with non-trivial endomorphisms have $\tau$ in some imaginary
quadratic field. In fact, it turns out that an elliptic curve $E$
has a non-trivial endomorphism if and only if
$\tau$ obeys a quadratic equation \quadreq\ with integer coefficients.
In this case,
$E$ is said to have complex multiplication (or to be of CM-type).
Summarizing, the endomorphism ring of a general elliptic curve
can be one of the following:
\eqn\ende{{\rm End} (E) = \cases{ \Z, & no CM \cr
\Z + \Z a \tau, & CM-type, $\tau = {-b + \sqrt{D} \over 2a}$} }
%
Thus, complex multiplication gives another way to characterize
elliptic curves with such `special' values of $\tau$.

There is a close relation between imaginary quadratic $\tau$,
which obey \quadreq, and binary quadratic forms:
$$
\pmatrix{2a & b \cr b & 2c}
$$
In our discussion, such forms will be associated with
intersection form of lattices. Notice, the form above
naturally defines a two-dimensional even lattice.
However, this form is not unique.
Namely, for any $S \in SL(2,\Z)$,
the intersection form:
$$
\pmatrix{2a' & b' \cr b' & 2c'} = S \pmatrix{2a & b \cr b & 2c} S^{{\rm tr}}
$$
defines the same lattice.
The invariant associated with such
a lattice is the discriminant $D=b^2-4ac$.
For a given value of $D$, the equivalence classes
of the integral binary quadratic forms form a finite abelian group,
the so-called {\it class group} \refs{\Borevicbook, \PSbook, \Sloane}:
$$
Cl(D)
$$
The order of this group $\vert Cl(D) \vert = h(D)$
is called the {\it class number}.
It is naturally identified with the number of ideal classes, $h(K)$,
of the imaginary quadratic field $K = \IQ (\sqrt{D})$.

The last object we need to introduce is an {\it order}, $\CO_f$.
\eqn\cofdef{ \CO_f = \Z \oplus \Z [f a \tau ]}
Here, $a$ a leading coefficient in the quadratic
polynomial \quadreq\ for $\tau$, and $f$ is a positive
integer number, called a {\it conductor} of $\CO_f$.
We can view an order $\CO_f$ as a two-dimensional lattice
in the number field $K = \IQ (\sqrt{D})$, generated by $1$ and $fa \tau$.
The reason $\CO_f$ will appear in our discussion is that
any element $\varphi \in \CO_f$ obviously gives a complex
multiplication, {\it cf.} \cmmap.
Note also, that the endomorphism ring itself, ${\rm End}~(K)$
is also an order with\foot{This special order in $K$ is called
the {\it ring of integers}.} $f=1$.

%
%
%
%


\newsec{General $c=2$ RCFT Based on the Elliptic Curve $E$}

In this section we are going to study general $c=2$ conformal
field theory based on elliptic curve with arbitrary parameters
$\tau$ and $\rho$. We will show that CFT is rational if
(and only if) both $\tau$ and $\rho$ take values in the same
imaginary quadratic field. The discriminant of this field
gives the dimension of the chiral ring. We also show that
a condition for diagonal modular invariant implies a further
relation between $\tau$ and $\rho$. Namely, one has to be
a complex multiplication for the other.

\subsec{Momentum Lattices}

In general, the partition function of $c=2$ CFT based on elliptic
curve $E = \IC / (\Z \oplus \tau \Z)$:
\eqn\pfncn{Z (q, \bar q) = {1 \over \eta^2 \bar \eta^2}
\sum_{ (p, \bar p) \in \Gamma^{2,2}}
q^{\half p^2} \bar q^{\half \bar p^2} }
is given by the sum over even, self-dual momentum lattice:
\eqn\lattice{\pmatrix{p \cr \bar p} \in \Gamma^{2,2} =
{i \over \sqrt{2 \tau_2 \rho_2}}
\Z \pmatrix{1 \cr 1}
\oplus \Z \pmatrix{\bar \rho \cr \rho}
\oplus \Z \pmatrix{\tau \cr \tau}
\oplus \Z \pmatrix{\bar \rho \tau \cr \rho \tau} }
This lattice will be one of the central objects in our discussion,
and, as we shall see in a moment, properties of the CFT,
like rationality {\it etc.}, can be formulated and analyzed
in terms of momentum lattices. For this reason, it is convenient
to introduce a few more objects associated with the momentum
lattice $\Gamma^{2,2}$.

First, we can define the following sublattices in $\Gamma^{2,2}$.
Let $\Gamma_0$ be the lattice of left-moving momenta $p$ for
a fixed value of $\bar p$ (say, for $\bar p = 0$)
and a similar lattice $\tilde \Gamma_0$:
\eqn\lzerolattice{\eqalign{
\Gamma_0 & = \{~ p~~ \vert~~ \pmatrix{p \cr 0} \in \Gamma^{2,2}~ \} \cr
\tilde \Gamma_0 & = \{~\bar p ~~\vert~~ \pmatrix{0 \cr \bar p}
\in \Gamma^{2,2}~\}
}}
Since $\Gamma^{2,2}$ is even integer lattice, the same is true
about $\Gamma_0$ and $\tilde \Gamma_0$.
The rank of these lattices is not greater than 2,
and generically it is zero.

By simply forgetting the right (or left) momentum,
we can also define the following projections:
\eqn\lrlattice{\eqalign{
\Gamma_{L} & = \{~ p~~ \vert~~ \pmatrix{p \cr *} \in \Gamma^{2,2}~ \} \cr
\Gamma_{R} & = \{~\bar p ~~\vert~~ \pmatrix{* \cr \bar p} \in \Gamma^{2,2}~\}
}}
Both $\Gamma_L$ and $\Gamma_R$ can be characterized as sets
where $p$ (or $\bar p$) take their values.
In general, unlike \lzerolattice, $\Gamma_{L,R}$ are not lattices.
Of course, in some special cases it may happen that the rank
of $\Gamma_{L,R}$ is less than 4. As we shall see below,
these are precisely the occasions relevant to rational theories.

Note, from the above definitions we obtain straightforward relations:
\eqn\split{\eqalign{
\Gamma_0 & \subseteq \Gamma_L \cr
\tilde \Gamma_0 & \subseteq \Gamma_R \cr
\Gamma_0 \oplus \tilde \Gamma_0 & \subseteq \Gamma^{2,2}
}}
%

\subsec{A Criterion for Rationality}

There are various ways of defining rational CFT's.
In our examples associated with tori, it is convenient to
formulate the condition for rationality in terms of momentum
lattices: a CFT is rational if and only if the left momentum
lattice $\Gamma_0$ is a finite index sublattice in $\Gamma_L$.
In such cases, both $\Gamma_0$ and $\Gamma_L$ are rank two
sublattices in $\Gamma^{2,2}$. Moreover, it is easy to see
that they are dual lattices:
\eqn\latticerel{\Gamma_0 \cong \Gamma_L^*,
\quad \tilde \Gamma_0 \cong \Gamma_R^*}
Indeed, since $\Gamma^{2,2}$ is even, self-dual integer lattice,
for any vector $(q, \bar q) \in \Gamma^{2,2}$ and
a given vector $(p,0) \in \Gamma_0$ we have a pairing:
$$
(p,0) \cdot (q, \bar q) = pq \in \Z
$$
Therefore, any vector in $\Gamma_0$ also belongs to $\Gamma_L^*$.
Conversely, to show $\Gamma_L^* \subseteq \Gamma_0$ let us take
a vector $p \in \Gamma_L^*$. Then, using the above equation
and self-duality of $\Gamma^{2,2}$, one finds that $(p,0)$
is a vector in $\Gamma^{2,2}$. Hence, $\Gamma_L^* \cong \Gamma_0$.
Similar arguments give the second isomorphism in \latticerel.

Therefore, we conclude that the study of rational conformal
field theories based on the elliptic curve $E$ is related
to the study of integer even two-dimensional lattices.
In particular, it gives a classification of such RCFT's.
We will come back to this later.

Now, let us discuss the geometric properties of the elliptic
curve corresponding to rational conformal field theory.
Suppose that a CFT associated with $E$ is rational,
{\it i.e.} $\Gamma_0$ is a finite index sublattice in $\Gamma_L$.
Using the explicit expression \lattice\ for the right
momenta $\bar p$, we find that the elements of $\Gamma_0$
correspond to integer numbers $(m_1, m_2, n_1, n_2) \in \Z^4$,
which obey two independent linear relations \Moore:
\eqn\gammaconstr{\eqalign{
m_1 + m_2 \rho + n_1 \tau + n_2 \tau \rho & = 0 \cr
m_1' + m_2' \rho + n_1' \tau + n_2' \tau \rho & = 0
}}
If we solve, for example, for $\rho$ from the second equation
and substitute the result to the first equation, we find
a quadratic equation for $\tau$, with integer
coefficients\foot{Alternatively, we could use the fact that any
three momentum vectors $p \in \Gamma_L$ (or $\bar p \in \Gamma_R$)
satisfy a linear relation over $\IQ$,
since $\Gamma_0$ is a $\Z \oplus \Z$-module.}:
\eqn\quadr{a \tau^2 + b \tau + c = 0, \quad gcd(a,b,c)=1}
with discriminant $D = b^2 - 4ac$.
Since imaginary part of $\tau$ has to be strictly positive,
{\it cf.} \physconstr, we conclude that $\tau$ has to belong
to imaginary quadratic number field $K = \IQ(\sqrt{D})$:
\eqn\tausol{\tau = {-b + \sqrt{D} \over 2a}}

If we now substitute this $\tau$ back into \gammaconstr,
we find that $\rho$ is linear in $\sqrt{D}$ over $\IQ$.
Hence, both $\tau$ and $\rho$ are elements in $K$.
In order to show that the converse is also true,
one can take $\tau, \rho \in K$, and construct,
for example, $\tau$ and $\tau \rho$ in terms of $1$ and $\rho$.
Since both $\tau$ and $\rho$ are assumed to be linear
functions of $\sqrt{D}$ with rational coefficients
(with $\tau_2 > 0$ and $\rho_2 > 0$), one can always
write $\tau$ and $\tau \rho$ as linear functions of
$\rho$ with rational coefficients.
Multiplying the resulting relations by a suitable integer,
one finds two equations of the form \gammaconstr\
with integer coefficients, where $n_2=0$ and $n_1'=0$.
Therefore, by construction these relations are independent
and define a lattice $\Gamma_0$.

Summarizing, we obtain an effective criterion for rationality
of $c=2$ conformal field theory based on the elliptic curve $E$:

\centerline{RCFT $\iff$ $\tau, \rho \in \IQ(\sqrt{D})$}
In other words, we found that in order for CFT to be rational,
both the target space torus and its dual should have complex
multiplication relative to the same quadratic imaginary field.

\subsec{Dimension Of The Chiral Ring And Verlinde Algebra}

As we found in the previous subsection, rational conformal
field theories based on the elliptic curve $E$ are
naturally attached to even integer lattices.
Specifically, let $v_i$, $i=1,2$, be the generators
of the momentum lattice $\Gamma_0$.
Since the intersection form is even, we can write it as:
\eqn\vvintersect{ v_i \cdot v_j = f \pmatrix{2a & b \cr b & 2c}}
for some integer numbers $a$, $b$, $c$, and $f$, such that $gcd(a,b,c)=1$.

By definition, the dual lattice $\Gamma_L = \Gamma_0^*$ is generated
by vectors $v_i^*$ with intersection form:
\eqn\dualintersect{
v_i^* \cdot v_j^* = {1 \over f D} \pmatrix{-2a & b \cr b & -2c}
}
where $D=b^2 - 4ac$ is the discriminant.
It is clear from \vvintersect\ and \dualintersect\ that
the dimension of the chiral ring, given by the index
$[\Gamma_L : \Gamma_0 ]$, is equal to:
$$
[\Gamma_L : \Gamma_0 ] = f^2 \vert D \vert
$$
Notice, that the right hand side is expressed
in terms of invariant quantities.

Furthermore, the Verlinde algebra is the group algebra of:
\eqn\discrgroup{D(\Gamma_0) = \Gamma_0^* / \Gamma_0}
In mathematics literature this group is usually called
the {\it discriminant group} \refs{\Nikulin,\Dolgachev},
see also \Sloane.
It is a finite abelian group of order $f^2 \vert D \vert$.
Since $\Gamma_0$ is a lattice of rank two, in general,
the discriminant group is a product of two cyclic groups,
in agreement with what we found in the specific examples
of tori with extra symmetries, {\it cf.} \verlindeguess.
Specifically, $D(\Gamma_0)$ is generated by two elements,
$g$ and $h$, such that:
$$
g^{2af} h^{bf} = 1  \quad {\rm and} \quad g^{bf} h^{2cf} =1
$$
The structure of this group depends in a crucial way on
the arithmetic. Specifically,
$$
D(\Gamma_0) = \cases{\Z_f \times \Z_{fD}, & $D=1$ mod 4 \cr
\Z_{2f} \times \Z_{2fD'}, & $D=0$ mod 4 ($D' = D/4$), $b \ne 0$ \cr
\Z_{2fa} \times \Z_{2fc}, & $b = 0$}
$$

This gives a general characterization of the Verlinde algebra in
the rational conformal field theory based on the elliptic curve $E$.

\subsec{Diagonal Modular Invariants}

Now we turn to the main problem, namely, analysis of
the conditions under which the partition function \pfncn\
takes the diagonal form \diagonal.
As we explained in the previous subsections, curve $E$
associated with a rational CFT has complex multiplication.
However, the ring ${\rm End} (E)$ itself did not enter
our discussion so far. Here, we show that
RCFT has a diagonal modular invariant iff either $\rho$ or
$\tau$ (or modular transformations thereof) belong to ${\rm End} (E)$,
i.e. iff $\rho$ is a complex multiplication for a given $\tau$,
up to discrete symmetries \symmetrygroup.

Diagonal modular invariant essentially implies identification
of left and right momentum lattices. Namely,
given an even integer lattice $\Gamma_0$,
one can canonically reconstruct the whole momentum lattice
$\Gamma^{2,2}$, which is even and self-dual.
Specifically, we take two copies of $\Gamma_L = \Gamma_0^*$:
\eqn\latticefroml{ \Gamma^{2,2} = (\Gamma_L , \Gamma_L') }
with the equivalence relation:
\eqn\latticerdef{\Gamma_L - \Gamma_L' = \Gamma_0}
to see that the lattice \latticefroml\ constructed in this
way is even, let us take a vector $(p , \bar p) \in \Gamma^{2,2}$.
By the equivalence relation \latticerdef\ we have $\bar p = p + v$,
where $v \in \Gamma_0$, and $p \in \Gamma_0^*$. Therefore,
\eqn\pevencheck{\eqalign{
(p , \bar p) \cdot (p , \bar p)
& = p^2 - \bar p^2
= (p + \bar p) (p - \bar p) = \cr
& = - v (2 p + v) = - 2 vp - v^2
}}
Since $\Gamma_0$ is taken even, both terms here are even and
the claim follows.
Furthermore, to show that $\Gamma^{2,2}$
constructed in \latticefroml\ is self-dual,
one can provide a basis,
$(p_i, \bar p_i) = \{ (v_i^*, v_i^*) , (v_j, 0) \}$,
where $v_j$ are the generators of $\Gamma_0$,
and $v_i^* \in \Gamma_0^*$ are the dual generators.
Then, the bilinear form looks like
$$
(p_i , \bar p_i) \cdot (p_j , \bar p_j) =
\pmatrix{0 & {\bf 1} \cr {\bf 1} & * }
$$
The determinant of this matrix is clearly equal to 1.
Therefore, $\Gamma^{2,2}$ is self-dual.

Now we want to compare a general lattice of
the form \dualintersect\ with the momentum lattice \lattice.
Up to discrete symmetries \symmetrygroup\
we can choose $\tau$ and $\rho$, such
that the left momentum lattice is generated by
(vectors proportional to) $1$ and $\tau$:
$$
\Gamma_0^* = \Gamma_L
= \{ {i \over \sqrt{2 \tau_2 \rho_2 }}, {i \tau \over \sqrt{2 \tau_2 \rho_2 }}\}
$$
It is easy to compute the intersection form of this lattice:
$$
v_i^* \cdot v_j^* = \pmatrix{ {1 \over 2 \tau_2 \rho_2 } &
{\tau_1 \over 2 \tau_2 \rho_2 } \cr {\tau_1 \over 2 \tau_2 \rho_2 } &
{ \vert \tau \vert^2 \over 2 \tau_2 \rho_2 } }
$$
Since $\Gamma_0$ has to be an even integer lattice,
one should be able to write this intersection as \dualintersect.
Comparing the individual entries, we find that in diagonal RCFT
$\tau$ and $\rho$ look like (up to discrete symmetries \symmetrygroup):
\eqn\taurhosol{\tau = {-b + \sqrt{D} \over 2a}, \quad \rho = fa \tau}
Hence, $\rho$ is a complex multiplication for
an imaginary quadratic $\tau$.
More precisely, $\rho$ should be associated with a generator
of an order in the imaginary quadratic field $K = \IQ (\sqrt{D})$:
$$
\rho \in \CO_f
$$
with the conductor $f$.

It is easy to verify that the converse is also true.
Namely, given $\tau$ and $\rho$ of the form \taurhosol,
the corresponding modular invariant is diagonal.
Indeed, substituting \taurhosol\ in the formulas
for the momentum vectors \lattice, we find that
the momentum lattice $\Gamma^{2,2}$ is of the form \latticefroml.
Specifically, we have:
%
%
\eqn\sa{\eqalign{
-i \sqrt{2 a f} \tau_2 ~p
& = m_1 + m_2 \bar \rho + n_1 \tau + n_2 \bar \rho \tau = \cr
& = (m_1 + fc n_2) + (n_1 - afm_2 + bf n_2) \tau }}
It immediately follows that:
\eqn\gammalgen{
\Gamma_L = \{ (p,*)^T \in \Gamma^{2,2} \}
= {i \over \sqrt{2 a f} \tau_2 }
( \Z \oplus \Z \tau ) }

In order to find elements of $\Gamma_0$, we have to solve $\bar p=0$.
For the right momenta we find:
\eqn\sb{\eqalign{
-i \sqrt{2 a f} \tau_2 ~\bar p
& = m_1 + m_2 \rho + n_1 \tau + n_2 \rho \tau = \cr
& = (m_1 - fc n_2) + (n_1 + afm_2 - bf n_2) \tau }}
Hence, $\bar p =0$ gives two conditions:
$$
\cases{m_1 = fc n_2 & \cr n_1 = - afm_2 + bf n_2 & }
$$
Substituting this into \sa\ yields:
$$
p \vert_{\bar p =0} = {i f \over \sqrt{2 a f} \tau_2}
\Big( (2a \tau + b)(-m_2) + (2c + b \tau)n_2 \Big)
\in \Gamma_0
$$
Therefore, we explicitly constructed $\Gamma_0$:
\eqn\gammanotgen{
\Gamma_0 = \{ (p,0)^T \in \Gamma^{2,2} \}
= {i f \over \sqrt{2 a f} \tau_2 }
( \Z [2a \tau + b] \oplus \Z [2c + b \tau ] ) }
The resulting lattices $\Gamma_L$, $\Gamma_0$, and $\Gamma^{2,2}$
are related as in \latticefroml\ -- \latticerdef, so that the
corresponding CFT is diagonal. It is straightforward to check
this directly computing the partition function \pfncn\ for these
momentum lattices.
As a result, one finds a diagonal modular invariant:
$$
Z(q, \bar q) = \sum_{\omega \in \CI}
\chi_{\omega} (q) \bar \chi_{\omega} (\bar q)
$$
where the exponent $\omega \in \CI$ that labels representations
of the chiral algebra can be identified with left
momenta, {\it cf.} \discrgroup:
\eqn\indset{\omega \in \Gamma_L / \Gamma_0}
and the characters $\chi_{\omega} (q)$ have the form:
\eqn\character{\chi_{\omega} (q) =
{1 \over \eta^2} \sum_{v \in \Gamma_0} q^{\half (v + \omega)^2} }
%


Finally, we can combine all the results in this section to conclude that
diagonal $c=2$ rational conformal field theories
are classified by the following data:

$i)$ discriminant $D=0,1$ mod 4
(a negative integer, such that $(-D)$ is square-free);

$ii)$ conductor $f$ (a positive integer);

$iii)$ an element of the class group $Cl(D)$.

\noindent
In terms of this data, the CFT has chiral ring of dimension
$\vert D \vert f^2$.

Suppose, for example, we want to know how many diagonal RCFT's
have the chiral ring of dimension 163. Using the above results,
we can immediately answer this question. Since $h(163)=1$, the answer
is surprisingly simple: there is a unique RCFT with this property.
On the other hand, if we asked a similar question about RCFT's
with chiral ring of dimension 159, we would find $h(159)=10$
such theories.
This simple example illustrates sporadic pattern of RCFT's,
which nevertheless can be completely explained by the methods
of number theory.

Note, it also follows from our analysis that the set of
$c=2$ rational conformal field theories on tori is dense.

\subsec{A Digression: The First Main Theorem of Complex Multiplication}

In this subsection (which can be skipped, especially in a single reading)
we digress on another remarkable property of elliptic curve
with complex multiplication. Throughout the paper we mainly viewed
the elliptic curve as a quotient of $\IC$ by a two-dimensional lattice.
However, we could also view $E$ as an algebraic curve
defines, say, by a Weierstrass polynomial:
\eqn\eweierstr{E \colon y^2 = 4x^3 - g_2 x - g_3}
where $g_2$ and $g_3$ are related to the modular parameter $\tau$
via invariant $j$-function:
\eqn\jfncn{\eqalign{
j(\tau) & = {1728 g_2^3 \over g_2^3 - 27 g_3^2}
= {\theta_{E_8} (\tau)^3 \over \eta(\tau)^{24}} = \cr
& = q^{-1} + 744 + 196844 q + 21493760 q^2 + 864 299 970 q^3 + \ldots
}}
Notice, that the coefficients in power series expansion
of the $j$-function are integer numbers.

Even though $j(\tau)$ is a very non-trivial function of $\tau$, there
is a nice characterization of its values for $\tau \in \IQ(\sqrt{D})$,
known as the first main theorem of complex multiplication.
Namely, suppose elliptic curve $E$ has a complex multiplication,
{\it i.e.} $\tau$ obeys a quadratic equation \quadreq.
Then, $j(\tau)$ also obeys a polynomial equation with integer
coefficients of degree $h$
(where $h=h(D)$ is the class number of the field $\IQ (\sqrt{D})$):
\eqn\hpolynom{P (z) = z^{h} + a_1 z^{h-1} + \ldots + a_h =0,
\quad a_i \in \Z}
A solution to such equation is called an algebraic integer
\foot{The first main theorem of complex multiplication further
says that if $z = j (\tau)$ is one of these numbers,
then $K(j(\tau))$ is the maximal abelian extension of $K = \IQ (\sqrt{D})$,
with ${\rm Gal} (K(j(\tau))/K) = Cl({\rm End} (E))$ acting transitively
on the set of numbers $j(\tau)$ \refs{\PSbook,\Borevicbook}.}.
Therefore, $j$-invariant of elliptic curve $E$ with complex
multiplication is an algebraic integer and $E$ is naturally
defined over the number field $K(j(\tau))$.

Motivated by this nice result, one might expect that a proper
criterion for CFT to be rational should be formulated as
a condition on the algebraic variety to be defined over
the algebraic closure $\bar{\IQ}$, obtained from the field $\IQ$
by adjoining the roots of all irreducible polynomials like \hpolynom.
It is easy to see, however, that this criterion would be wrong.
Indeed, it would predict ``too many'' RCFT's. For example,
in the case of Calabi-Yau manifolds it would predict existence
of infinitely many points (which are dense) in the moduli space,
whereas in the later sections we will argue to the contrary.

More elementary is to see that the above criterion fails
already in the case of the elliptic curve $E$. Indeed,
in general, the converse of the first main theorem of
complex multiplication is not true, so it can not be
formulated as ``if and only if'' condition. On the other
hand, from the analysis of the previous sections, we know
that CFT is rational {\it if and only if} $E$ has complex
multiplication. This demonstrates that the right signature
of the rational CFT is complex multiplication, rather than
a possibility to define the target space variety over $\bar{\IQ}$.

\newsec{Geometric Interpretation of Cardy States}

In the previous sections we have established a relation
between RCFT data and arithmetic of the elliptic curve $E$.
Motivated by such a relation, one may wonder if it
can be extended to string theory, including D-branes
and other non-perturbative objects. Due to their geometric
nature, D-branes seem to be especially promising.
In the weak coupling limit they can be viewed as
submanifolds of $E$ of various (co)dimension.

{}From the CFT point of view, there are some `special' D0-branes,
which preserve the full chiral algebra. The corresponding
boundary states were explicitly constructed by Cardy \Cardy.
Therefore, one could ask: ``What is arithmetic/geometric
interpretation of the Cardy states?''
There are several natural candidates for the answer to this question.
For example, once we deal with arithmetic of $E$, one might
consider rational points of $E$, {\it i.e.} solutions of
\eweierstr\ with rational values of the coordinates $x,y \in \IQ$.
However, these can not correspond to Cardy states.
Indeed, the number of rational points on $E$ may be infinite,
whereas the number of Cardy states is always finite (and equal
to the dimension of the chiral algebra):
$$
\# ({\rm Cardy~states})~ = \vert D \vert f^2
$$
where $D$ is the discriminant of the quadratic polynomial
\quadr\ for $\tau$, and $f$ is the conductor.

In this section we study D-branes on elliptic
curve $E$ with diagonal modular invariant:
\eqn\diagonalzz{Z (q, \bar q)
= \sum_{j \in \CI} \chi_{j} (q)~ \bar \chi_{j} (\bar q)}
where the exponent $j$ can be identified with momentum, {\it cf.} \indset:
\eqn\jindset{j \in \Gamma_L / \Gamma_0}
and the characters have the form \character.
Following Cardy \Cardy, we show that there are
always $\vert D \vert f^2$ D0-branes in this theory,
which correspond to the regular points of $\Gamma_L / \Gamma_0$.
On the other hand, the number of D2- and D1-branes depends
on the arithmetic of the elliptic curve in a very interesting way.
For example,
$$
\# ({\rm D2-branes})~= \cases{1 & $Df$ odd \cr 4 & $Df$ even}
$$
Note, simple examples studied in section 3 agree with this
general result, see \dsquaretorus\ and \dsuthreetorus.
Below we explain these results in more detail.

Before we start, let us remind that Cardy states are linear
combinations of Ishibashi states \bstate:
\eqn\ishi{\vert B \rangle \rangle
= \exp \Big[ - \sum_{n=1}^{\infty} {1 \over n}
\alpha^{\mu}_{-n} R_{\mu \nu} \tilde \alpha^{\nu}_{-n} \Big]
\times \sum
\vert p, \bar p \rangle
}
which satisfy boundary conditions \boundarycond.
It is convenient to write this boundary condition in
the following form:
\eqn\boundarycondp{p = - R \bar p}
Since labels $j$ in \jindset\ are identified with momenta,
we can say that for a given $R$ the Ishibashi states are
labeled by:
$$
(j, - Rj), \quad j \in \Gamma_L / \Gamma_0
$$
However, since we assume that RCFT is diagonal \diagonalzz,
the Ishibashi state $(j, \bar j)$ appears in the closed string
spectrum only if $j = \bar j$:
$$
(j,j) = (j, - Rj)
$$
It follows that for a given gluing condition \boundarycondp\
the Ishibashi states are labeled by fixed points of $R$ in
the exponent set $\CI$:
\eqn\ishiset{j = - R j, \quad j \in \Gamma_L / \Gamma_0}
In particular, the number of solutions to this conditions
give us the number of the corresponding D$p$-branes.
In the following subsections we solve \ishiset\ for each value of $p$.
Note, that $j=0$ is always a solution to \ishiset\ for any allowed $R$,
{\it i.e.} there is always at least one corresponding D-brane.

\subsec{D0-branes}

In order to get a D0-brane, we have to impose Dirichlet
boundary conditions in both spatial directions, so that
$2 \times 2$ matrix $R$ must have two eigenvalues $-1$, {\it i.e.}:
$$
R = - {\bf 1}
$$

In this case, the boundary condition \boundarycondp\ has the
$$
p = \bar p
$$
It does not impose any further constraints on the momentum,
so that and $p \in \Gamma_L / \Gamma_0$ is a solution.
Hence, the number of D0-branes is given by the following
universal formula for all models:
\eqn\dzeronumber{\# ({\rm D0-branes})~= \vert D \vert f^2}
This number is the same as the dimension of the chiral ring,
and suggests interpretation of D0-branes as special
points on the torus $E$. Indeed, we can think of D0-branes
as regular points in the quotient $\Gamma_L / \Gamma_0$
or else, as preimages of a marked point on a torus $E$
under a specific complex multiplication
\foot{The complex multiplication $\hat \rho$ has a number of special properties.
First, note that it can be obtained by taking a derivative
of the defining quadratic polynomial for $\tau$:
$$
\hat \rho = f {d Q (\tau) \over d \tau},
\quad Q (\tau) = a \tau^2 + b \tau + c
$$

Another distinguished property of $\hat \rho$ is that it
corresponds to a complex multiplication whose
square is multiplication by an integer.
Moreover, it is the only complex element (up to integer multiples)
in the ${\rm End} (E)$ with this property.
Explicitly, the square of the element $2a\tau+b$ is:
$$
(2a\tau+b)^2 = 4a^2\tau^2+4ab\tau+b^2
= 4a(a\tau^2+b\tau)+b^2 = 4a(-c)+b^2
$$
which is the discriminant, $D$.} (see Figure 4):
$$
z \mapsto \hat \rho \cdot z
$$
where, {\it cf.} \taurhosol:
\eqn\rhohat{\hat \rho = f(2a \tau + b)}
Indeed, one can easily check that $\hat \rho \cdot 1 = f(2a \tau + b)$
and $\hat \rho \cdot \tau = - f(2c + b \tau)$ are the two
generators of the lattice $\Gamma_0$, {\it cf.} \gammanotgen.

\ifig\pic{D0-branes in rational CFT can be identified with
points (black dots) in the lattice $\langle 1, \tau \rangle$ modulo
$\langle \hat \rho, \hat \rho \tau \rangle$. In this figure
we illustrate this for a specific model where $\tau$
satisfies the quadratic equation: $\tau^2 + 2 \tau + 2 = 0$.
In this example, $a=1$, $b=2$, $c=2$, and $D=-4$.}
{\epsfxsize4.0in\epsfbox{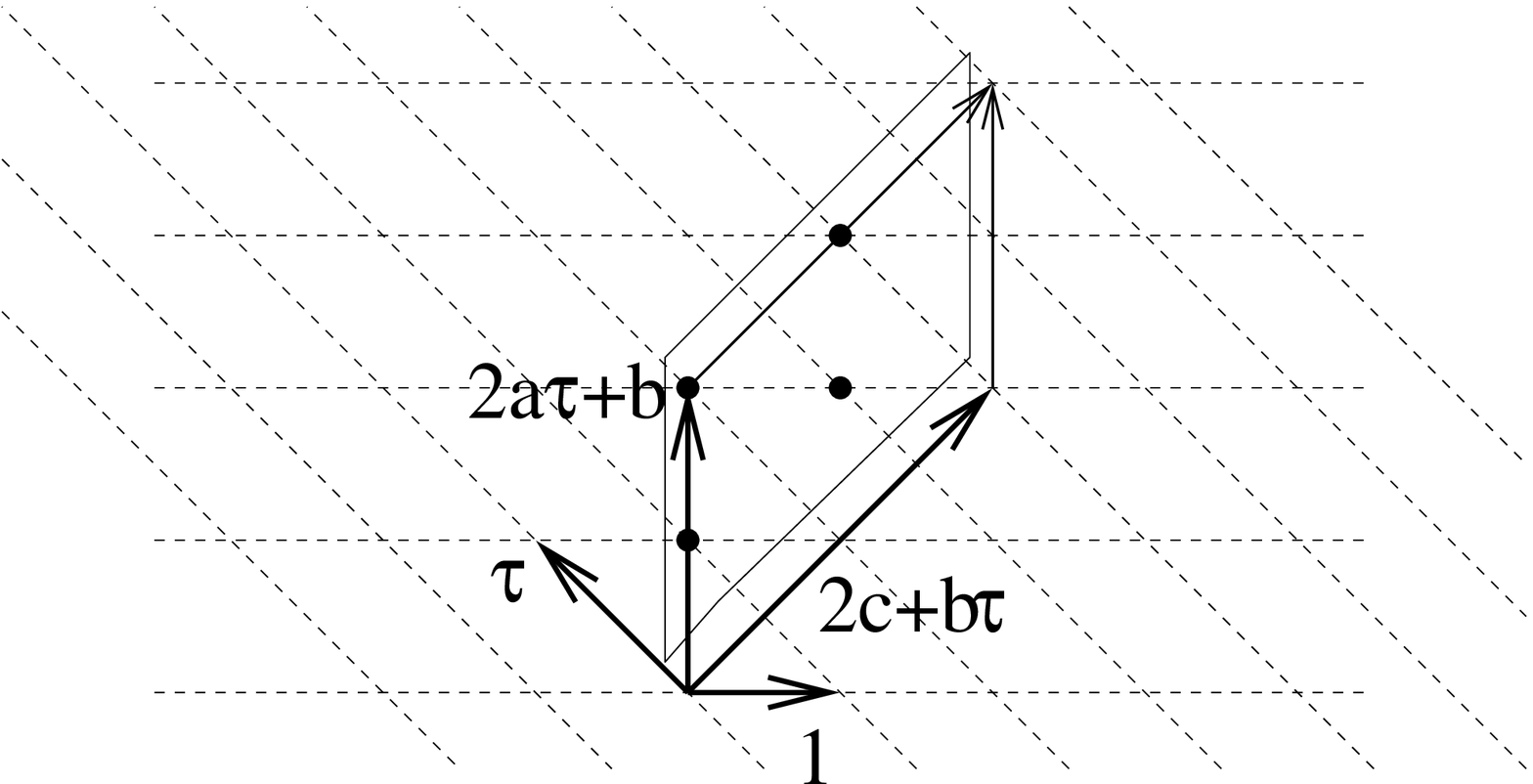}}

Geometrically, it is convenient to visualize the set
$\Gamma_L / \Gamma_0$ as a parallelogram with edges
$f(2a \tau + b)$ and $f(2c + b \tau)$ in the lattice $\Gamma_L$.

Note that $\hat \rho =f(2a\tau +b)$ is closely related
to the K\"ahler structure of the torus, which is $\rho =fa\tau$.
In fact, if $bf$ is even  $\hat \rho$ can be viewed as twice the
K\"ahler class (with a suitable shift of $\rho $ by $bf/2$.  Thus
roughly speaking the K\"ahler class defines the relevant endomorphism
of the torus which defines the Cardy states by its preimage.

\subsec{D2-branes}

The next simplest case is when we impose Neumann boundary
conditions in all the directions:
$$
R = {\bf 1}
$$
The involution $R$
inverts the exponent set $\CI = \Gamma_L / \Gamma_0$:
\eqn\dtwobc{R \quad : \quad p \mapsto - p}
and flips the parallelogram subtended by vectors
$f(2a \tau + b)$ and $f(2c + b \tau)$, as shown in Figure 5.

\ifig\pic{In the case of D2-brane boundary conditions,
the involution $R$ flips the parallelogram made by vectors
$f(2a \tau + b)$ and $f(2c + b \tau)$. The origin (black dot)
is always fixed under this involution. The other three
potential fixed points are denoted by empty circles.}
{\epsfxsize4.0in\epsfbox{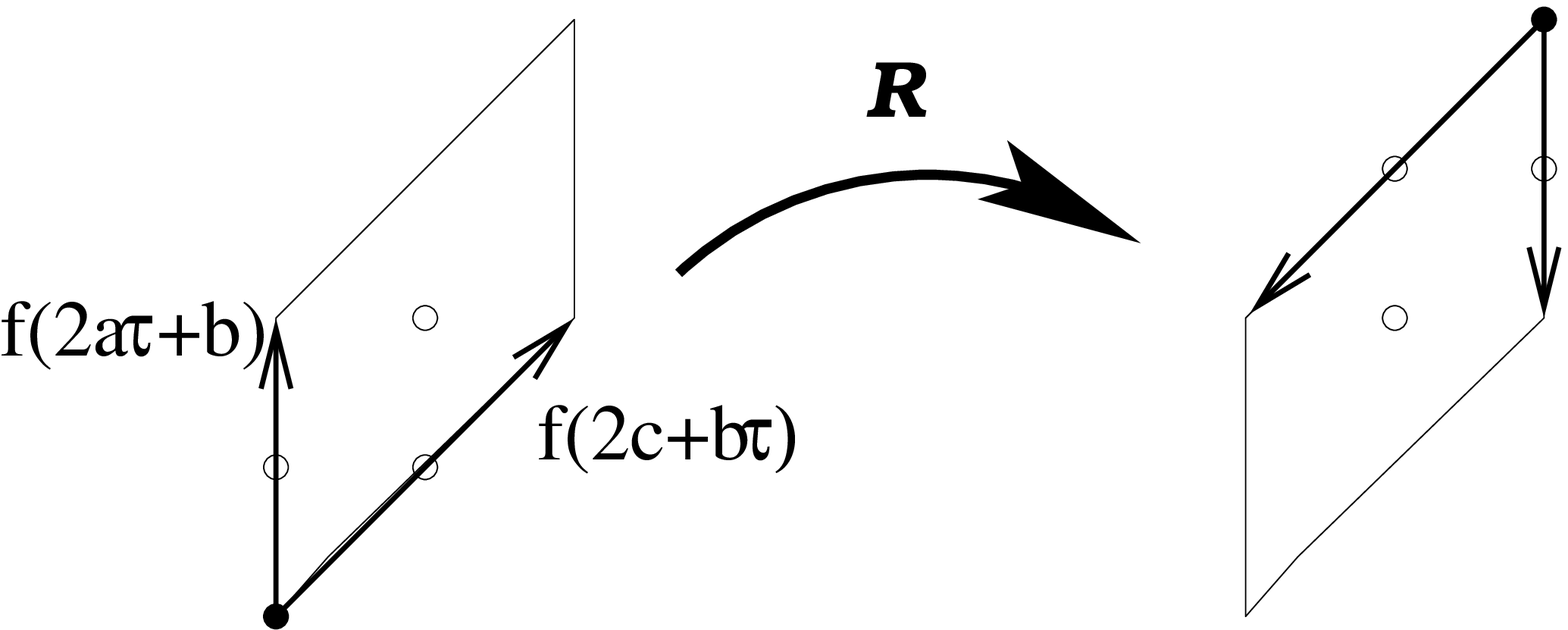}}

According to \ishiset, the Ishibashi states are in one-to-one
correspondence with the fixed points of this involution,
modulo the lattice $\Gamma_0$. Geometrically, it is clear
that there is either one or four fixed points on the parallelogram
(see Figure 5), depending on whether its edges have odd or even
coordinates in the lattice $\Gamma_L \cong \{ 1, \tau \}$.
Indeed, the origin is always a fixed point of $R$.
Let us see when there are extra fixed points.
The potential candidates are middle points on the edges of
the parallelogram and a point in the middle (denoted by empty
circles in Figure 5). The explicit coordinates of these points
in the lattice $\Gamma_L \cong \{ 1, \tau \}$ are the following:
$$
(fb/2,fa), \quad
(fc, fb/2), \quad
(fc + fb/2, fa + fb/2)
$$
It is clear that all of these points are in the lattice
$\Gamma_0$ if and only if both $fb$ is even. Hence, we
arrive to the following general result:
\eqn\dtwonumber{
\# ({\rm D2-branes})~= \cases{1 & $bf$ odd \cr 4 & $bf$ even} }
Note that in the case of $bf$ even, the four inequivalent
$D2$ branes differ by the value of $Z_2$ Wilson lines along the two cycles.

\subsec{D1-branes}

Finally we consider the case of D1-branes. In this case
we have one Neumann and one Dirichlet direction, which correspond
to $+1$ and $-1$ eigenvalues of $R$, respectively. Allowing for
D1-branes of arbitrary orientation, we can write the corresponding
involution $R$ as:
\eqn\donebc{R \quad : \quad p \mapsto \alpha p^*}
where $\alpha$ is some phase, $\vert \alpha \vert =1$.  Note
that this is an order 2 operation.
Since $p$ takes values in the lattice $\Gamma_L \cong \{ 1, \tau\}$,
the involution $R$ must respect this lattice. In particular
it should map a basis of $\Gamma_L$ into another basis:
\eqn\basismap{\eqalign{1 \mapsto & -A  - B\tau \cr
\tau \mapsto & C + D\tau }}
where $A$, $B$, $C$, $D$ are integer numbers, such that $AD-BC=1$.
Therefore, we get two conditions:
\eqn\abasismap{\eqalign{\alpha = &- A  - B \tau\cr
\alpha \bar \tau = & C  + D\tau }}
{}from which we can eliminate $\alpha$:
\eqn\donetau{-\bar \tau = {A + B \tau \over C  + D\tau }}
Simply put,
the last condition says that $-\bar \tau$
should be an involution of $PSL(2;\Z)$ acting on $\tau$, for otherwise
we wouldn't have any D1-branes. Then, for every $\tau$, which
satisfy a relation of the form \donetau, there might be
different involutions corresponding to different $\alpha$'s.
Therefore, it is natural to split the question in two parts,
and classify first all $\tau$ which solve \donetau, and
then classify all possible $\alpha$'s.

\ifig\pic{Two families of the solution for $\tau$ lie
either on a unit circle or on the imaginary line.
There are also two special cases, $\tau=i$ and
$\tau=\exp (2 \pi i / 3)$, represented by black dots.
A number near every point indicates the total number of
D1-brane boundary states in the corresponding RCFT.}
{\epsfxsize3.0in\epsfbox{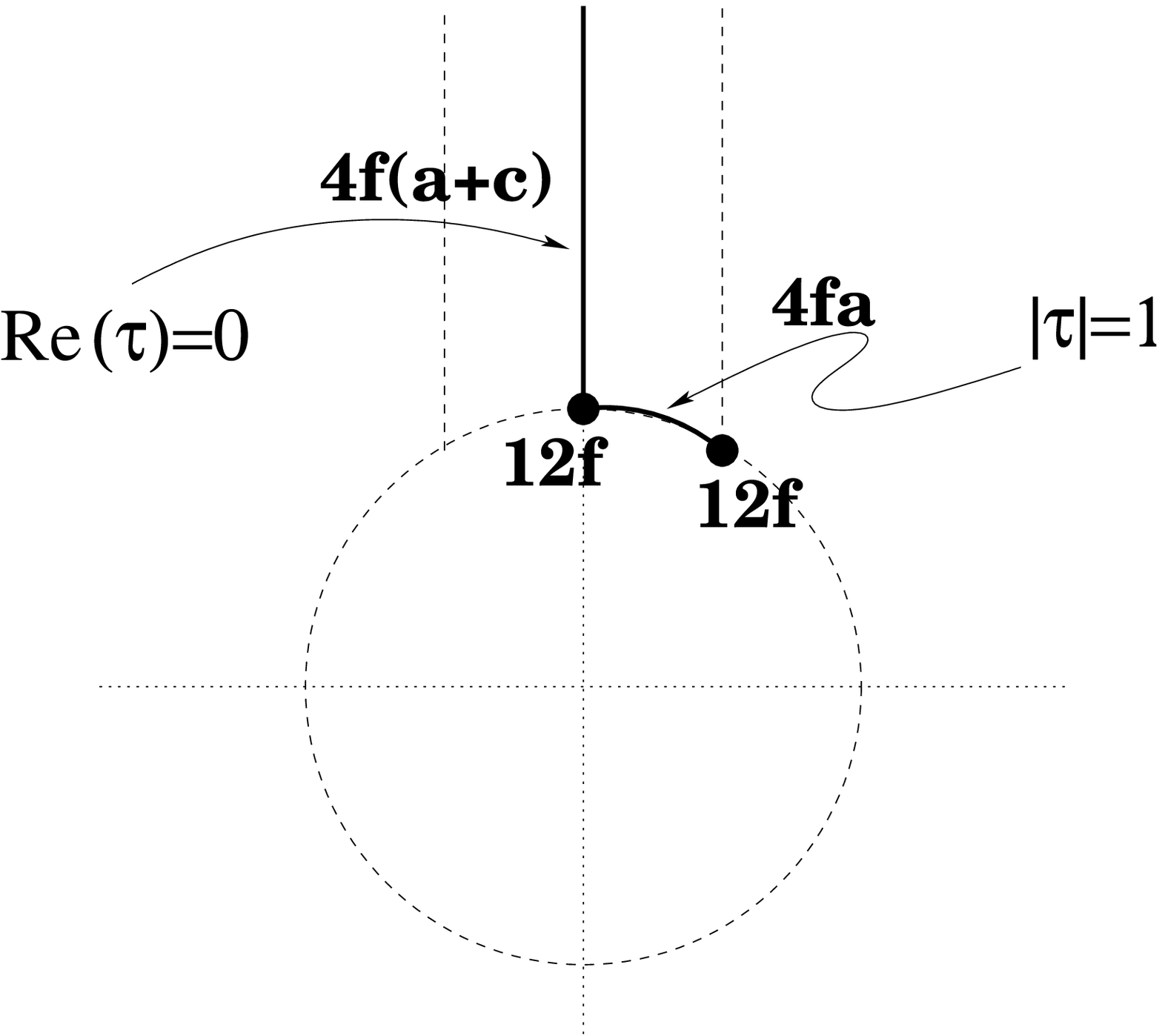}}

It is easy to see that the only $\tau$ (in the fundamental domain),
which solve \donetau, are $(i)$ either pure imaginary, $\tau_1=0$,
or $(ii)$ lie on a unit circle, $\vert \tau \vert =1$, see Figure 6.
To see that these are the only solutions, with no loss of generality
we can assume $\tau $ is in the fundamental domain of the upper half
plane.  We are looking for involutions of $SL(2,Z)$ acting
on it which give $-\overline \tau$.  On the other hand
$-\overline \tau$ is also in the fundamental domain
of the upper half plane.  This can only be consistent
with the notion of the fundamental domain if the involution
is $\pm $ the identity matrix, or it is $\pm S$ in which case it
maps the boundary of the fundamental domain to itself.  There
is one extra case corresponding to the involution $TST^{-1}$ which
fixes $\tau ={\rm exp}({2\pi i/3})$.  In the first
case we have $\tau=-{\overline \tau}$ which states that $\tau$ is pure
imaginary.  In the latter case it implies that $\tau$ is on the unit circle.

The two different families of the solutions for $\tau$
correspond to elliptic curves with different symmetries,
simple examples of which were discussed in section 3.
Let us now consider each of the two cases in turn:

\bigskip\noindent{\it (i) -- a product of two circles}

In the first case $b=0$ and the solutions to \donetau\ look like:
$$
\tau = \sqrt{ - {c \over a}}
$$
In fact, this is precisely the case discussed in section 3.1,
where the torus is a product of two circles:
$$
E = {\bf S}^1 \times {\bf S}^1
$$
All possible D1-branes in this case were classified
in \dsquaretorus, with $k_1=fa$ and $k_2=fc$:
\eqn\donenumberi{\# ({\rm D1-branes})~= 4 f(a + c)}
and come from $\alpha =1 $ or $\alpha =-1$.
They correspond to $2fa$ equally spaced $D1$ branes along one direction
and $2fc$ equally spaced $D1$ branes along the other direction of the torus.
Moreover each of these one branes can have a $\Z_2$ Wilson line on them.

\bigskip\noindent{\it (ii) -- a torus symmetric relative to the diagonal}

In this case $|\tau|=1$ (which implies
 $a=c$) and the torus $E$ has extra symmetry
corresponding to the reflections relative to the diagonals.
In particular $z\rightarrow \alpha {\overline z}$ is a symmetry
when $\alpha =\pm \tau$, which thus generically
yields two involutions $R$.  For the case of $\tau =i$ we have
4 involutions given by $z\rightarrow
i^k {\overline z}$.  For $\tau =e^{2\pi i/3}$
we have 6 involutions given by $z\rightarrow \omega^k {\overline z}$
where $\omega$ is a 6-th root of unity. The corresponding
fixed characters under this involution will correspond
to equally spaced D1 branes in the direction given by $\sqrt{\alpha}$.

The corresponding values of $\tau$ are:
$$
\tau = {- b + \sqrt{D} \over 2a}, \quad D = b^2 - 4a^2
$$

For the generic case, the total number of D1-branes, which  is just
the number of the fixed lattice points under the two reflections is
\eqn\donenumberi{\# ({\rm D1-branes})~= 4fa}
For the case $\tau =i$ the D1 branes making angles multiple of $\pi/4$.
For $D1$ branes in the direction $2n \pi/4$ ($n=0,1$)
we have $2f$ equally
spaced branes each of which can have an extra $Z_2$ Wilson line.
For $D1$ branes in the directions $\pi/4,3\pi/4$ we have $2f$ equally
spaced branes (all branes passing through the origin) without any
extra possibility of Wilson lines.  The $D1$ branes corresponding
to the latter angles do not come from tensoring Ishibashi
states of the two decoupled circles, but rather it corresponds
to using the extra $Z_2$ exchanging the two circles.
For $\tau =e^{2\pi i/3}$ we have $D1$ branes which make
angles of $2\pi n/ 12$, where $n=0,...,5$.  One can check that for $n$ even
there are $3f$ equally spaced ones and for $n$ odd there are $f$ equally
spaced ones. In both special cases ($\tau=i$ and $\tau=\exp (2 \pi i / 3)$)
the total number of D1-branes is $12f$.

We have thus seen that the classification of $Dp$ branes for $p>0$
is much more sporadic than those for the $D0$ branes.  This
is to be expected because the $D0$ branes are precisely the ones naturally
picked out by the diagonal modular invariant.


\newsec{RCFT and Higher Dimensional Calabi-Yau }

So far we have talked about CFT's based on the simplest
Calabi-Yau sigma model, namely the target being $T^2$. More
precisely, we focused only on the bosonic sigma model, but
in this case the incorporation of fermions does not modify our
discussion, as the fermion partition function is independent of
$T^2$ moduli. It is natural then to raise the question of
RCFT's corresponding to supersymmetric sigma models propagating
on higher dimensional Calabi-Yau manifolds.

Unfortunately not much is known about the exact solutions in such
cases, and we only have existence proof for such CFT's.  The only
general classes we know are tensor products of minimal $\CN=2$ supersymmetric
conformal theories (Gepner models) and toroidal orbifolds.  It turns
out that {\it both} classes are RCFT's or have moduli for which
RCFT's appear in a dense subspace. It is the existence of this
class of examples which motivates the belief that CFT's are ``exactly
solvable'' if they are rational or near one.
Consider quintic threefold, for example.  There is only one point in
its moduli of K\"ahler and complex deformation where the CFT is exactly
known and that is the RCFT corresponding to the Gepner point.
One wonders whether there are other points on the moduli space where
they are rational and therefore, perhaps solvable.

In this section we propose a criterion for rationality
of conformal theory on Calabi-Yau manifolds which agrees with
all the known examples of rational points discussed above
(see \Wendland\ for a further evidence in the case of toroidal
conformal field theories).
However, given mathematical conjectures
{\it a la} Andr\'e and Oort \refs{\Andre,\Oort}
our proposal for rationality suggests that RCFT's
{\it are not dense} in the generic case of Calabi-Yau sigma models!

Our criterion for rationality is motivated by generalization of
the notion of complex multiplication to higher dimensional varieties.
Indeed, it was pointed out to us by Kazhdan and Mazur that there
already exists a suitable notion of complex
multiplication\foot{Mathematically, it says that manifold $M$
has complex multiplication when its Hodge group, ${\rm Hg} (M)$,
is commutative.} for higher dimensional varieties introduced
in 1969 by D.~Mumford \Mumford.
In particular, complex multiplication was studied in the context of
K3 surface by Piateckii-Shapiro and Shafarevich \PShaf, and more generally,
in the context of Calabi-Yau manifolds by Borcea \Borcea.
The idea is rather simple: 
One first defines what it means for an abelian variety
(i.e. complex tori) to admit complex multiplication.
Then one asks if the variation of the Hodge structure of
the Calabi-Yau $M$ and its mirror $W$,
whose period matrices lead to a pair of associated abelian
varieties admit complex multiplication (of a `compatible' type).


\subsec{Complex Multiplication for Complex Tori}

Consider a complex $n$-dimensional torus.
$$
T^{2n} \cong \IC^n / \Z^{2n}
$$
This is defined by identifications
$$z_i\sim z_i+\delta_{ij}\qquad z_i\sim z_i+\CT_{ij}$$
where $\CT$ is an $n\times n$ complex symmetric
period matrix.  Then we say that the torus admits complex
multiplication if there exists a non-trivial endomorphism
\eqn\amap{z \mapsto A z}
which implies that
$$
A=M+N\CT
$$
$$
\CT A=M'+N' \CT
$$
for some integer matrices $M,N,M',N'$.
In other words we have a second order matrix equation
$$\CT (M+N\CT)=M'+N'\CT \quad \Rightarrow $$
\eqn\baseq{\CT N \CT + \CT M-N' \CT -M'=0.}
Moreover one requires that $N$ has rank $n$.\foot{
This rules out examples like $M=E_{CM} \times E'$,
where $E_{CM}$ is an elliptic curve with complex
multiplication and $E'$ is another arbitrary elliptic
curve without CM.}

\subsec{Calabi-Yau and the Intermediate Jacobian}

The notion of considering mid dimensional cohomology
elements and integrating over a mid dimensional integral
lattice of cycles to define an abelian variety is well known.
For example for the case of a genus $g$ Riemann surface
with a symplectic pairing of 1-cycles:
\eqn\abpairing{(A_i,B_j)=\delta_{ij}\quad (A_i,A_j)=(B_i,B_j)=0}
one considers $g$ holomorphic 1-forms $\omega_i$ normalized
relative to the $A$-cycles
$$\int_{A_j}\omega_i=\delta_{ij}$$
and defines the Jacobian by
\eqn\tmatrix{\int_{B_j}\omega_i=\CT_{ij}.}
Similar idea works for arbitrary complex varieties and in
particular for Calabi-Yau 3-folds.  In the case of Calabi-Yau
threefold $\CT_{ij}$ can be identified with the complex
torus defining the coupling constants of the associated
$U(1)^n$ gauge fields and is related to the prepotential $\CF$
(in the homogeneous coordinates) by
$$\CT_{ij}=\partial_i\partial_j \CF$$
We say that the Calabi-Yau admits complex multiplication if
the corresponding intermediate Jacobian associated with
$\CT$ admits complex multiplication.

\subsec{A Criterion for RCFT for Calabi-Yau Sigma Models}

A Calabi-Yau sigma model is completely characterized
by its complex and K\"ahler moduli.  Since the notion
of complex multiplication is natural only for complex
structure of the variety, it is natural to
associate to a given Calabi-Yau $M$ with a given complex
and K\"ahler structure, a mirror pair of Calabi-Yau $(M,W)$
with fixed complex structures (where we have traded the K\"ahler
structure of the original Calabi-Yau with complex moduli
of its mirror).  This is effectively how we studied
the case of elliptic curve, by viewing $\tau$ and $\rho$
as defining pairs of elliptic curves.  We propose the following
criterion of the Calabi-Yau sigma model to
correspond to a RCFT:

\bigskip

{\it Sigma model on Calabi-Yau corresponding
to the pair $(M,W)$ is RCFT if and only if
$M$ and $W$ admit complex multiplication
over the same number field.}

\bigskip

For example, for a Calabi-Yau threefold $M$ satisfying
complex multiplication one gets an equation \baseq\ of order
two in the $(h^{2,1}(M)+1) \times (h^{2,1}(M)+1)$ matrix $\CT$.
In this case, it has been shown by Borcea \Borcea\ that
existence of complex multiplication is equivalent to
the condition that elements of the endomorphism matrix $A$
generate imaginary number field $K$:
\eqn\cyfieldk{K \cong {\rm End} (H^3(M,\IQ)) \otimes \IQ}
of degree:
$$
[K:\IQ] = 2(h^{2,1}(M)+1)
$$
On the other hand,
the mirror $W$ admitting complex multiplication gives elements
which are in an algebraic number field of degree
$2(h^{2,1}(W)+1) = 2(h^{1,1}(M)+1)$,
and the criterion we are imposing for RCFT is that
they are elements of the same number field\foot{Clearly,
the degree of this number field is bounded by
${\rm min} (2(h^{2,1}(M)+1), 2(h^{1,1}(M)+1))$.}.

\subsec{Application of the Criterion}

To check the criterion, we have to make sure it agrees
with the known cases of RCFT's for Calabi-Yau sigma models.
Indeed it does.  Toroidal orbifolds corresponding
to RCFT's obviously admit complex
multiplication inherited from the fact that the underlying
torus admits complex multiplication (extending our discussion
from the elliptic case --- the simplest case being orbifolds
of the product of elliptic curves).  Much more non-trivial
are the Gepner points, corresponding to Fermat polynomials.
It is also known that these also do admit complex
multiplication \refs{\Shioda,\Deligne,\MDeligne,\Borcea}.
Below we show how this works for the quintic threefold
with one complex moduli. This is already impressive evidence
for the criterion we have proposed for rationality.

We now wish to study how frequently one would encounter
rational conformal theory in the moduli of a given Calabi-Yau
sigma model, assuming the criterion we have proposed holds.
To get a feel for this, consider Riemann surfaces.
As discussed above we can identify with it an associated Jacobian.
However the moduli space of genus $g$ curves is $3g-3$ complex dimensional
whereas the moduli space of the abelian varieties of dimension $g$
has dimension $g(g+1)/2$. Thus for $g>4$ the Riemann surfaces are not
dense in moduli of the corresponding tori. The Schottky problem is
to identify which abelian varieties can arise for Riemann surfaces.

Similarly, one could ask which Riemann surfaces
admit complex multiplication.  Even though there is a dense
set of point in the moduli of complex structure of the tori
admitting complex multiplication this may not hold true for the
measure zero subspace of it corresponding to those coming
from Riemann surfaces.  Unlikely as this sounds, indeed there
is evidence and a standing mathematical conjecture by
Coleman \Coleman\ (see also \JNoot\ for recent developments)
that for sufficiently large $g$ there are only a {\it finite}
number of Riemann surfaces admitting CM!
Indeed a similar conjecture exists for arbitrary
varieties\foot{Mathematically, a basic version of this conjecture
is known as Andr\'e-Oort conjecture \refs{\Andre,\Oort},
and we thank F.~Oort and B.~Mazur for explaining to us
the general philosophy behind it. Roughly, Andr\'e-Oort conjecture
says that in order for a (sub)family of algebraic varieties to
contain a dense set of CM-points, the corresponding moduli
space has to be ``Shimura (sub)variety''. For example, the moduli
spaces of elliptic curves and $K3$ surfaces are of this type, however
the moduli space of a Calabi-Yau manifold in general is not.}
and it is believed that the number of CM points are dense
only if the relevant moduli space itself is of the form $G/H$
(i.e. a submoduli of the full toroidal moduli defined by some
linear algebraic constraint).
In particular, in the case of complex tori (of complex
dimension $n$) and $K3$ surfaces this conjecture predicts
dense set of CM/RCFT points. Indeed, in both cases
the moduli space turns out to be a coset space:
\eqn\tmspace{SO(2n,2n;\Z) \backslash SO(2n,2n) / SO(2n) \times SO(2n)}
and
$$
SO(20,4;\Z) \backslash SO(20,4) / SO(20) \times SO(4)
$$
respectively. On the other hand, for the case of the one parameter
family of quintic three-folds, complex multiplication
is conjectured to occur at most at {\it finite} number of points.
It would be very interesting to test this conjecture as
it seems to be at odds with the common lore for RCFT's.
This of course might be a blessing in disguise as
it seems to point to the existence of some finite number
of interesting points on the moduli of Calabi-Yau compactifications.
These may end up being interesting points when the moduli
of Calabi-Yau manifolds get frozen by some mechanism.

\subsec{The Example of Fermat Quintic}

Finally, a non-trivial test of our criterion can be obtained
by considering a one-dimensional family of quintic three-folds
$$
M: \quad
z_1^5 + z_2^5 + z_3^5 + z_4^5 + z_5^5 - 5 \psi z_1 z_2 z_3 z_4 z_5 =0
$$
At the Fermat point, $\psi=0$, the corresponding sigma-model
becomes rational, namely it is the $(k=3)^5$ Gepner model.
On the other hand, Calabi-Yau manifold $M$ has
complex multiplication at $\psi=0$ \refs{\Borcea,\Shioda,\Deligne}.
This is related to the fact that the automorphism group is bigger for
the Fermat quintic than for any other generic member in this family.
Moreover, $\psi=0$ is the only known non-trivial CM-point
in the whole moduli space of $M$.
In this sense, there is the same amount of physical and
mathematical data on this question, which therefore
provides at least one non-trivial check of our proposal.

In order to see explicitly that the Fermat quintic has
sufficiently many holomorphic endomorphisms (and, therefore,
admits complex multiplication) let us evaluate the period
matrix \tmatrix\ at $\psi=0$.
In a particular basis of $A$ and $B$ cycles \abpairing,
the standard calculation gives \refs{\Candelas,\IHV}:
$$
\CT = \pmatrix{\a - 1 & \a + \a^3 \cr \a + \a^3 & - \a^4}
$$
where $\a$ is a (non-trivial) 5-th root of unity, $\a^5=1$.
Note, that $\a$ is a solution to the degree 4 polynomial
with integer coefficients:
$$
x^4 + x^3 + x^2 + x + 1 =0
$$

It is straightforward to check that the matrix $\CT$
satisfies the quadratic matrix equation of the form \baseq:
$$
\CT N \CT + \CT M-N' \CT -M'=0
$$
where
$$
N = \pmatrix{1 & -1 \cr 0 & 1}, \quad
M = \pmatrix{0 & 0 \cr 0 & 0}, \quad
N' = \pmatrix{-1 & 0 \cr -1 & 0}, \quad
M' = \pmatrix{-1 & 0 \cr -1 & -1}, \quad
$$
The corresponding endomorphism is given by the matrix:
$$
A = \pmatrix{\a-1 & \a + \a^3 \cr 1 + \a + \a^3 & -\a^4}
$$
Notice, that elements of $\CT$ and $A$ take values in
a degree 4 number field $K$, {\it cf.} \cyfieldk:
$$
K = \IQ (\a)
$$
which can be obtained from the field of rational numbers, $\IQ$,
by adjoining the fifth root of unity.
Since in the present example $h^{2,1}(M)=1$,
this is in complete agreement with the general
formula for the degree, $[K:\IQ] = 2(h^{2,1}+1)$.

\bigskip
\centerline{\bf Acknowledgments}
We would like to thank D.~Kazhdan and B.~Mazur
for many illuminating discussions on complex multiplication.
We are also grateful to J.~de Jong, J.~Maldacena,
K.~Oguiso, H.~Ooguri, F.~Oort, A.~Recknagel, S.~Shenker,
F.~Rodriguez-Villegas, and E.~Witten for valuable discussions.
This research was partially conducted during the period S.G. served as
a Clay Mathematics Institute Long-Term Prize Fellow. The work of
S.G.~is also supported in part by grant RFBR No. 01-02-17488,
and the Russian President's grant No. 00-15-99296.
The work of C.V. is supported in part by
NSF grants PHY-9802709 and DMS 0074329.

\listrefs
\end